\documentclass[12pt]{article}
\usepackage{amsmath,amssymb,amsfonts,amsthm,bbm,cancel,wasysym}
\usepackage{epsfig,graphics,graphicx,epstopdf}
\usepackage{rotating}
\usepackage{colordvi,color}
\usepackage{array,booktabs,colortbl,colordvi,multirow}
\usepackage{setspace}
\usepackage{enumerate}
\usepackage{enumitem}
\usepackage{jheppub}
\usepackage{float}
\usepackage{changepage}
\usepackage{ulem}
\restylefloat{table}

\newcommand{\ctoprule}{\toprule[0.5mm]}
\newcommand{\cbottomrule}{\bottomrule[0.5mm]}

\newcommand{\crowcolor}{\rowcolor[rgb]{0.9,0.9,0.9}}

\renewcommand{\arraystretch}{2.0}

\newcommand{\be}{\begin{equation}}
\newcommand{\ee}{\end{equation}}
\newcommand{\bea}{\begin{eqnarray}}  
\newcommand{\eea}{\end{eqnarray}}

\begin{document}


\begin{flushright}
UG-FT-306/13, 
CAFPE-176/13 \\
%
\end{flushright}
%
%
%
%

\title{LHC bounds on Lepton Number Violation \\ mediated by doubly and singly-charged scalars}

\date{\today}

\author{Francisco\ del \'Aguila \footnote{E-mail: faguila@ugr.es}}
\author{and Mikael Chala \footnote{E-mail: faguila@ugr.es}}
\affiliation{CAFPE and Dpto. de F\'{\i}sica Te\'orica y del Cosmos \\
Universidad de Granada, E-18071, Granada, Spain}

\abstract{
The only possible doubly-charged scalar decays into two Standard Model particles are into pairs 
of same-sign charged leptons, H$^{\pm\pm}\rightarrow {l}^\pm {l}^\pm,  {l} = e, \mu, \tau$, 
or gauge bosons, H$^{\pm\pm}\rightarrow W^\pm W^\pm$; being necessary the observation of both 
to assert the violation of lepton number. 
However, present ATLAS and CMS limits on doubly-charged scalar production are obtained under 
specific assumptions on its branching fractions into dileptons only. 
Although they can be extended to include decays into dibosons and lepton number violating 
processes. 
Moreover, the production rates also depend on the type of electroweak multiplet H$^{\pm\pm}$ 
belongs to. 
We classify the possible alternatives and provide the Feynman rules and codes for generating the 
corresponding signals for pair and associated doubly-charged scalar production, including the 
leading contribution from the $s$-channel exchange of electroweak gauge bosons as well as the 
vector-boson fusion corrections. 
Then, using the same analysis criteria as the LHC collaborations we estimate the limits on the H$^{\pm\pm}$ 
mass as a function of the electroweak multiplet it belongs to, and obtain the bounds on the lepton number 
violating processes 
$pp\rightarrow {\rm H}^{\pm\pm}{\rm H}^{\mp\mp} \rightarrow {\ell}^\pm {\ell}^\pm W^\mp W^\mp$ and 
$pp\rightarrow {\rm H}^{\pm\pm}{\rm H}^{\mp} \rightarrow {\ell}^\pm {\ell}^\pm W^\mp Z$, 
$\ell = e, \mu$,  
implied by the ATLAS and CMS doubly-charged scalar searches.   
}

\maketitle




\section{Introduction}
\label{Intro}
\renewcommand{\arraystretch}{0.5}

No departure from the Standard Model (SM) predictions has been observed at the LHC, yet; 
being the properties of the recently discovered Higgs boson apparently SM-like, too 
\cite{Aad:2012tfa, Chatrchyan:2012ufa,Giardino:2013bma}. 
This was also preferred by electroweak (EW) precision data \cite{Barbieri:2000gf,delAguila:2011zs} 
(see for previous fits \cite{Goebel:2010ux} 
and for an up-date \cite{Baak:2012kk,deBlas:2013gla}), 
thus making the discovery of new physics (NP) in the first LHC run unlikely.
What leaves the well-established neutrino masses as the only clear signal of NP 
beyond the SM, so far \cite{Beringer:1900zz}, if we obviate the cosmological evidence 
of dark matter. 

Neutrinos are massless in the SM because they have no right-handed (RH) counterparts, 
$\nu_{R i}$, to form Dirac masses and lepton number (LN) is an accidental symmetry 
protecting them to acquire Majorana masses \cite{Mohapatra:1998rq}. 
Hence, in order to describe neutrino masses we have to add new degrees of 
freedom to the SM: either RH neutrinos with the corresponding Yukawa couplings 
giving Dirac masses to neutrinos after EW symmetry breaking (EWSB): 
${\cal L}^{Y}_{m_\nu} = - y_{ij} \overline{L_{L i}} \nu_{R j} \tilde\phi + {\rm h.c.} 
\rightarrow - y_{ij} v \overline{\nu_{Li}} \nu_{Rj} + {\rm h.c.}$,
with 
$v = \langle\phi^0\rangle \sim 174$ GeV the Higgs vacuum expectation value (VEV) 
{\footnote{$L_L = (\nu_L, l_L)$ and $\phi = (\phi^+, \phi^0)$ are the SM lepton and Higgs 
doublets, respectively, with $\tilde{\phi} = i\sigma_2 \phi^*$ and $\sigma_2$ the second 
Pauli matrix. 
We write down column doublets in a row for convenience, when no confusion is expected.}}; 
or new (heavy) fields, which in particular may be also 
$\nu_{R i}$, with couplings violating LN explicitly or spontaneously and generating 
Majorana masses for the SM neutrinos at some given order in perturbation theory.  
In this case, upon integration of the heavy modes, the model is described at low energy 
by an effective Lagrangian with extra higher-order operators, the one of lowest dimension 
being the Weinberg operator \cite{Weinberg:1979sa}, 
$\mathcal{O}^{(5)} = (\overline{L^c_L}\tilde{\phi}^*)(\tilde{\phi}^\dagger L_L)$, 
parametrizing the neutrino Majorana masses 
{\footnote{$L^c_L = (\nu^c_L, l^c_L)$ is the SM lepton doublet with charge-conjugated fields, 
$\psi^c_L = (\psi_L)^c = C\overline{\psi_L}^T$; 
analogously $\psi^c_R = (\psi_R)^c = C\overline{\psi_R}^T$.}}: 
${\cal L}^{(5)}_{m_\nu} = - c^{(5)}_{ij} {\cal O}^{(5)}_{ij} / \Lambda + {\rm h.c.} 
\rightarrow - (c^{(5)}_{ij} v^2 / \Lambda) \overline{\nu^c_{Li}} \nu_{Lj} + {\rm h.c.}$, 
with $\Lambda$ the scale of NP. 

In this second case there are no new light degrees of freedom to start with; 
the simplest realizations being characterized at low energy 
by the very tiny LN violation (LNV) induced by the neutrino Majorana masses, 
$(m_{\nu})_{ij} = 2\ c^{(5)}_{ij} v^2 / \Lambda \sim 0.1$ eV. Whose measurement 
is the purpose of the next generation of neutrinoless double $\beta$ decay 
experiments \cite{Furry:1939qr} (for recent reviews see \cite{Vergados:2002pv,Avignone:2007fu}) 
\footnote{Although the leading contribution to this process may come from other 
(higher-order) operators in more elaborated models 
\cite{de Gouvea:2007xp,delAguila:2011gr,delAguila:2012nu,Gustafsson:2012vj,Franceschini:2013aha}}. 
However, the relevant question in the LHC era is if LNV is at the LHC reach. 
This is to ask if there are new particles with masses $\Lambda \sim$ TeV 
(and then $c^{(5)}_{ij} \sim 10^{-11}$) with observable LNV signatures 
\cite{Keung:1983uu} 
\footnote{If the observed baryon asymmetry of the universe originates from 
leptogenesis, LNV must be at work at some energy, too \cite{Fukugita:1986hr}. 
(See for recent reviews \cite{Davidson:2008bu}.)}. 
There is a wide literature dealing with the 
simplest realizations of this scenario, which are referred to as 
{\it see-saw} mechanisms of type I, II and III and obtained 
extending the SM 
with RH neutrinos 
\cite{Minkowski:1977sc}, 
a scalar triplet 
\cite{Schechter:1980gr} 
and vector-like fermion triplets \cite{Foot:1988aq}, respectively. 
In order to assert the violation of LN at the LHC it is enough to observe final 
states with non-zero LN, for the LN of the initial state ($pp$) vanishes. 
What in practice means observing events with an excess of leptons, 
or anti-leptons. 
Among the three {\it see-saw} mechanisms, the {\it see-saw} of type II gives the 
cleanest signal because doubly-charged scalars can decay into pairs of same-sign 
charged leptons, $\Delta^{\pm\pm} \rightarrow l^\pm l^\pm$, 
which accumulate around the doubly-charged scalar mass and allow 
for a very efficient search 
\cite{Hektor:2007uu,delAguila:2008cj}; 
which is not the case for heavy fermions because they decay into an 
odd number of light fermions 
($\geq 3$, if we exclude decays into Higgs bosons 
decaying in turn into two photons) 
\cite{Han:2006ip,Franceschini:2008pz}, 
as required by rotational invariance. 
As a matter of fact, CMS \cite{Chatrchyan:2012ya} and ATLAS \cite{ATLAS:2012hi} 
have already set stringent limits on this process, 
excluding doubly-charged scalar masses $m_{\Delta^{\pm\pm}}$ ranging 
from 200 to 400 GeV, depending on the assumptions on the branching ratios 
into same-sign dileptons; 
although they have not reported on the corresponding limits for LNV. 
In order to obtain the latter, one must look for events with same-sign charged 
lepton pairs plus EW gauge bosons, including the production of 
$pp \rightarrow {\rm H}^{\pm\pm} {\rm H}^{\mp\mp} \rightarrow l^\pm l^\pm W^\mp W^\mp$ 
and 
$pp \rightarrow {\rm H}^{\pm\pm} {\rm H}^{\mp} \rightarrow l^\pm l^\pm W^\mp Z$ 
\footnote{In the text generic scalar multiplets are denoted by H and 
their doubly-charged component by H$^{\pm\pm}$. We use  $\Delta$ 
when only referring to the scalar triplet. 
While charged leptons are denoted by $l (\ell)$ when tau leptons 
are (not) included.}, 
what can be done with the same sampling, four and three isolated leptons 
plus possibly missing transverse momentum, as we discuss below. 

First, however, several general comments are worth to emphasize again:\hfill\break
({\it i}) LNV is minuscule, and hence at the LHC the production of LNV particles 
must be very suppressed or their decay very slow. As in the former case 
these would not be observable, they must transform non-trivially under the 
SM gauge symmetry and hence be produced with EW strength 
\footnote{We assume that the new fields do not carry color because 
we search for LNV particles which mainly manifest as dileptonic resonances.}. 
(Obviously, singlets can be produced through mixing with non-singlet 
states but this mechanism is in general suppressed by the corresponding mixing angles. 
An example is heavy neutrino production through mixing with SM leptons 
\cite{Han:2006ip}, which is suppressed 
because the corresponding mixing angles are 
bounded to be small by EW precision data (EWPD) \cite{delAguila:2008pw} 
\footnote{Even if vector boson fusion contributions are large 
\cite{Dev:2013wba}, EWPD including LHC data on the SM Higgs 
further reduce the limits on lepton mixing \cite{deBlas:2013gla} and 
hence, the LHC potential for heavy neutrino detection \cite{Han:2006ip}. 
In any case, LHC direct limits on heavy neutrino production 
provide independent evidence and restrict the allowed range of 
heavy neutrino masses \cite{Chatrchyan:2012fla,ATLAS:2012yoa}, 
which are indirectly not accessible to lowest order in the expansion in the 
small lepton mixing. 
On the other hand, the LHC reach for heavy neutrino detection 
can be much larger in the presence of new interactions. 
In particular, if parity is restored \cite{LR} and the new charged gauge boson 
$W'$ has a mass of several TeV, heavy Majorana neutrinos and hence 
LNV events can be observed up to neutrino masses near the $W'$ mass 
\cite{WR_N}. In fact, CMS has already set significant bounds 
on this process \cite{CMS:2012zv}.}.)\hfill\break
({\it ii}) Thus, LN must be violated in the decays of the new heavy particles, 
what requires that they have at least two dominant channels with different LN. 
(Majorana fermions are charge self-conjugated and hence if they decay 
into a final state with non-zero LN, they do also decay into the charge-conjugated 
state with opposite LN.)\hfill\break
({\it iii}) We restrict ourselves to SM extensions with LNV scalars 
because, as stressed above, scalar signatures allow for a more efficient particle reconstruction. 
Moreover, although the discovery of the Higgs boson proves the fundamental 
character of the SM scalar sector at low energy, this remains the less 
known sector of the model. 
In summary, the experimental observation of neutrino masses together with the 
outstanding LHC performance make the search for LNV scalars (eventually 
contributing to neutrino masses) especially timely. 
In the following we shall extend the {\it see-saw} of type II, which is mediated by an 
$SU(2)_L$ scalar triplet with hypercharge $Y=1$, 
$\Delta = (\Delta^{++},\Delta^{+},\Delta^{0})$, to allow for scalar 
multiplets H with arbitrary isospin $T$ and hypercharge $Y$ but with a doubly-charged 
component ${\rm H}^{++}$ coupling to a pair of same-sign charged leptons. 
This alone fixes the scalar LN equal to $- 2$ but does not stand for LNV. 
In order to violate LN the scalars must also decay into SM boson pairs 
and then to final states with vanishing LN. Indeed,  
the only other possible two-body decay into SM particles of the doubly-charged 
scalar is into two $W$ bosons, 
in the scalar triplet case $\Delta^{\pm\pm} \rightarrow W^\pm W^\pm$. 
This decay does require the LN breaking by the (small) non-zero $\Delta^0$ VEV. 
In general, once LN is broken, doubly-charged scalars ${\rm H}^{\pm\pm}$ 
with non-vanishing LN will also decay into $W$ pairs at some order in 
perturbation theory; whether 
its neutral partner ${\rm H}^0$, if it exists, gets a (small) non-zero VEV, or through mixing 
with other (heavier) scalar multiplets with diboson couplings. 
The consideration of doubly-charged scalar decays into dileptons and dibosons  
on the same footing in order to search for (bound) LNV at the LHC also generalizes 
previous phenomenological studies. 

In simple models the region of parameter space where 
the doubly-charged scalar branching ratio into 
two same-sign leptons is comparable to the branching 
ratio into gauge bosons is small. In general, 
one of the two couplings is larger than the other and therefore 
the corresponding decay dominates. However, both decays can naturally 
have a similar rate in more elaborated models 
\cite{Schechter:1981bd,delAguila:2011gr,delAguila:2012nu,Gustafsson:2012vj,Babu:2009aq}
\footnote{\label{mixing}
We also assume that the mass splitting between 
the different components of the multiplet is small and hence the 
mixing with heavier scalar multiplets and with the SM Higgs. 
Otherwise, cascade decays within the multiplet (of electroweak 
strength) would be overwhelming 
\cite{Grifols:1989xe,Perez:2008ha}. 
Anyway, if ${\rm H}^{\pm\pm}$ mainly decays 
into ${\rm H}^{\pm} W^{\pm *}$, their subsequent leptonic decays 
also involve neutrinos, then making more difficult (less efficient) 
to reconstruct the doubly-charged scalar \cite{Aoki:2011pz}. 
Moreover, the final fermions are softer and do not exhibit the resonant 
behavior in the same-sign dilepton channel. 
We will not further consider this scenario in the following.}.\hfill\break
({\it iv}) We extend the SM with an extra TeV scalar multiplet at a time, 
neglecting possible mixing effects with other heavier scalar multiplets except 
to allow for the decay of the TeV scalar multiplet into gauge bosons. 
The only models we shall work out in detail are those with scalar multiplets with 
components of charge 2 at most, which are those of smaller isospin, too 
\footnote{\label{triply}
Multiplets with components with larger charges 
also have other striking signatures, for instance 
${\rm H}^{+++}\rightarrow {\rm H}^{++ *}W^{+ *}\rightarrow l^+l^+l^+\nu$, 
but with less energetic charged leptons in the final state \cite{Babu:2009aq,Hisano:2013sn}. 
In any case doubly-charged scalars are in general pair 
and associated produced with a comparable 
cross-section, decaying besides into 
same-sign dileptons and dibosons as assumed 
here and hence with harder charged leptons in the final state, which make 
easier (more efficient) the doubly-charged scalar reconstruction.}.\hfill\break
({\it v}) We will not discuss flavor constraints either because they are model dependent at a large extent. Thus, although in the {\it see-saw} of type II neutrino masses are proportional to the corresponding doubly-charged scalar decays \cite{Hektor:2007uu,delAguila:2008cj,Raidal:2008jk}, in general they are not closely related and more elaborated models can accommodate both independently of their specific values. At any rate, along this paper our approach to LHC searches will be mainly phenomenological and hence largely model independent. \hfill\break
({\it vi}) Once LNV is observed the question will be which its origin is. 
In the case of doubly-charged scalar production under consideration one 
would like to determine the type of multiplet the doubly-charged scalar 
belongs to. This can be done sampling appropriately the events 
with four and three isolated leptons, as has been proposed in 
\cite{delAguila:2013yaa,delAguila:2013hla}. 

In next section 
we characterize the scalar multiplets with doubly-charged components 
decaying into pairs of same-sign charged leptons, {\it i.e.}, 
the possible isospin and hypercharge multiplet assignments. 
In general, the larger their isospin is, the higher the dimension 
of the operators parametrizing the heavy scalar decay and hence smaller their decay rate. 
Their gauge interactions are detailed in Section \ref{sec:gauge}, where 
we work out the corresponding Feynman rules. 
Both sections are more technical and can be skipped if the reader is only 
interested in the phenomenological implications. 
The production mechanisms are discussed in Section \ref{sec:prod}. 
The dominant mechanism for doubly-charged scalar pair and associated 
production is through the $s-$channel exchange of EW gauge bosons. 
Vector-boson fusion contributions staying below 10 \% for the scalar masses 
of interest. 
The software implementation for Monte Carlo simulations is described in 
Section \ref{sec:Simulation}, being available upon request. 
Section \ref{sec:analyses} contains the analyses mimicking those performed by 
the ATLAS and CMS experiments; 
and we extend them to estimate the bounds on LNV in Section \ref{sec:Efficiencies}. 
In particular, we provide a table with an estimate of the efficiencies for the reconstruction 
of the different decay modes, which allows to derive the corresponding limits 
on doubly-charged scalar production for any set of branching ratios and hence model. 
We conclude in Section \ref{sec:concl}. 
In Appendices \ref{effectiveproduction}, \ref{sec:SimulationAndAnalyses} and \ref{statistics} 
we gather further technical details on effective operators for doubly-charged scalar 
production at hadron colliders, the Monte Carlo implementation for 
doubly-charged scalar pair and associated production and the applied statistics, 
respectively.

%
%
\section{Which kind of new physics are we looking for ?}
\label{sec:class}

We want to search for scalar resonances that may decay into 
a pair of same-sign charged leptons, ${\rm H}^{\pm\pm} \rightarrow l^\pm l^\pm$, 
and be eventually at the LHC reach. 
This means to classify the EW multiplets H which the corresponding 
doubly-charged scalars can belong to. 
In general, no matter what $SU(2)_L\times U(1)_Y$ multiplet including H$^{\pm\pm}$ 
is considered, one can always write down gauge invariant effective operators 
giving rise to these decays after EWSB \cite{delAguila:2013yaa}. 
In fact, this can be done for any of the three lepton bilinears with non-vanishing LN 
available in the SM: $\overline{L_L^c}L_L$, $\overline{l_R^c}l_R$ and $\overline{L_L^c}l_R$, 
the three of them containing the product of two same-sign charged leptons $l^-l^-$. 
Although we can restrict ourselves to the first two combinations because the  
operators involving the third one are not independent of those built with the first two: 
the third combination $\overline{L_L^c}l_R$ requires a $\gamma_\mu$ insertion 
because of the fermions' chirality, and hence the presence of a covariant derivative 
to ensure the operator is Lorentz invariant; then using integration by parts and the 
equations of motion, the corresponding operators can be seen to be equivalent to the
ones involving $\overline{L_L^c}L_L$ and $\overline{l_R^c}l_R$. 

In practice we assume that there is a more fundamental theory reducing at lower energy 
to the SM plus an extra scalar multiplet H near the TeV scale with LN$=-2$ and a 
doubly-charged component H$^{++}$ 
\footnote{Doubly-charged fermions and vector-bosons have been also 
considered but in other context \cite{Alloul:2013raa}.}. 
Hence, its isospin $T$ and hypercharge $Y$ 
must fulfill 
\begin{equation}
\label{chargerelation}
T^{\rm H} \geq | T_3^{{\rm H}^{++}} = 2 -Y^{\rm H} | ;
\end{equation}
and for any pair of isospin and hypercharge assignments satisfying this relation 
there is a tower of gauge invariant operators involving H, any of the two 
bilinears with LN = 2, $\overline{L_L^c}L_L$ or $\overline{l_R^c}l_R$, and an 
increasing number of Higgs doublets $\phi (\tilde{\phi})$, with vanishing LN.  
This reflects the fact that any $SU(2)_L\times U(1)_Y$ representation 
satisfying Eq. (\ref{chargerelation}) can be obtained from the Clebsch-Gordan 
series of the product of a large enough number of fundamental representations 
$\phi (\tilde{\phi})$, with $T =1/2$ and $Y=1/2 (-1/2)$. 
In particular, one can correlate the operators involving 
$\overline{\tilde L_L} \tau^a L_L$ 
\footnote{Where ${\tilde L_L} = i\sigma_2L_L^c$ and $\tau^a$ are the Pauli matrices in the spherical basis, 
$A^{+1} = -\frac{1}{\sqrt 2} (A_1-iA_2), A^0 = A_3, A^{-1} = \frac{1}{\sqrt 2} (A_1+iA_2)$,  
times the Clebsch-Gordan coefficients ${\rm C}^{1\times 1\rightarrow 0}_{a,-a}$, up to a 
global factor and sign: 
$\tau^{\pm1} = \pm (\sigma_1\mp i \sigma_2)/2,~ \tau^0 = \sigma_3/\sqrt{2}$.}, 
with $T = 1$ and $Y = -1$, to those involving $\overline{l_R^c}l_R$, with 
$T = 0$ and $Y = -2$, contracting the former with $\phi^\dagger \tau^{-a} \tilde{\phi}$; 
and vice-versa multiplying by ${\tilde\phi}^\dagger \tau^{a} {\phi}$.
However, for any given H only the operators of lowest dimension in general matter 
because they are the ones formally giving the largest contributions to the dileptonic 
H$^{++}$ decays after EWSB. 

\begin{figure}[t]
\begin{center}
\begin{tabular}{cc}
\includegraphics[width=0.2\columnwidth]{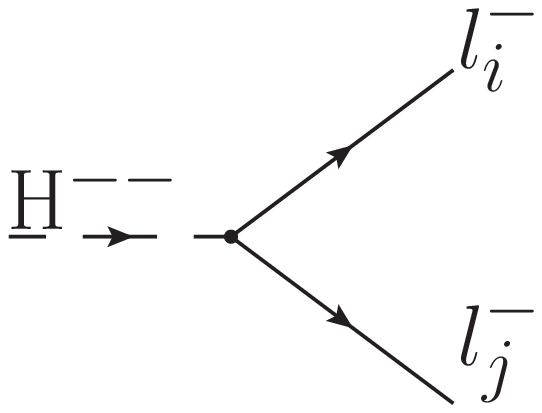} \raisebox{0.8cm}
{$2i\left[\alpha_{ij}^{L *} P_L+\alpha_{ij}^{R *} P_R\right]$} & 
\includegraphics[width=0.2\columnwidth]{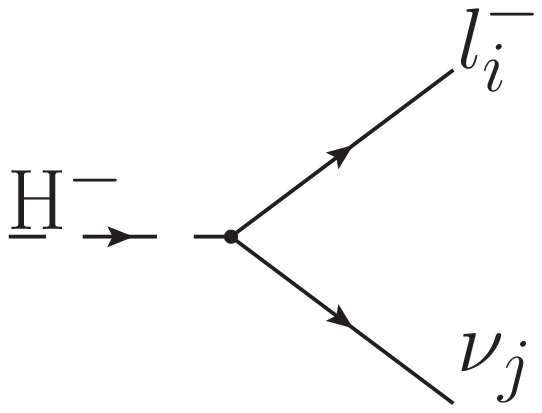} \raisebox{0.8cm}{$2i\beta^*_{ij}P_L$}\\
\end{tabular}
\caption{Feynman diagrams and rules for Yukawa interactions. The arrows indicate the LN flow.}
\label{table:YukawaRules}
\end{center}
\end{figure}

For illustration purposes in the following we restrict ourselves to scalar multiplets with 
at most doubly-charged components: 
\begin{equation}
\label{chargelessequaltwo}
 T^{\rm H} = T_3^{{\rm H}^{++}} \leq 2 .
\end{equation}
This stands for an $SU(2)_L$ singlet $\kappa^{++}$ with hypercharge 2 
\cite{delAguila:2011gr,Gustafsson:2012vj,Babu:1988ki}, 
a doublet $\chi=(\chi^{++},\chi^+)$ with $Y=3/2$ 
\cite{Gunion:1996pq}, 
a triplet $\Delta=(\Delta^{++},\Delta^+,\Delta^0)$ with $Y=1$ mediator 
of the {\it see-saw} of type II \cite{Schechter:1980gr}, 
a quadruplet $\Sigma=(\Sigma^{++},\Sigma^+,\Sigma^0,\Sigma'^-)$ with $Y=1/2$ 
\cite{Ren:2011mh}, 
and a quintuplet which we will assume to be real  
$\Omega=(\Omega^{++},\Omega^+,\Omega^0,\Omega^-,\Omega^{--})$ 
with $Y=0$. 
The lowest order gauge invariant operators coupling the doubly-charged 
component of these multiplets to a pair of same-sign charged leptons  
after EWSB are of dimension 
4 for $\kappa$ and $\Delta$, 5 for $\chi$ and $\Sigma$, and 6 for $\Omega$, 
respectively \cite{delAguila:2013yaa}:
\begin{equation}\label{op:singlet}
\mathcal{O_\kappa} = \overline{l_R^c}l_R\kappa; \qquad
\mathcal{O}_{\Delta}  = (\overline{\tilde L_{L}} \tau^a L_{L}) M^\Delta_{ab}\Delta^{b}, 
\quad {\rm with} \;\; a, b = 1,0,-1; \nonumber
\end{equation}
\begin{equation}\label{op:doublet}
\mathcal{O}_\chi^{(1)} = 
\overline{l_R^c} l_R (\tilde{\phi}^\dagger\chi);
\qquad \mathcal{O}_\chi^{(2)} = 
(\overline{\tilde L_{L}} \tau^a L_{L}) 
M^\chi_{ab}(\phi^\dagger \tau^b\chi), \quad {\rm with} \;\; a, b = 1,0,-1; \nonumber
\end{equation}
\begin{equation}\label{op:quadruplet}
\mathcal{O}_{\Sigma}  = 
(\overline{\tilde L_{L}} \tau^a L_{L})M^\Sigma_{c,ab}\phi^b \Sigma^{c}, 
\quad {\rm with} \;\; a = 1,0,-1, \;\; b = \pm \frac{1}{2} , \;\; c =\frac{3}{2}, \frac{1}{2}, -\frac{1}{2}, -\frac{3}{2} ; \nonumber
\end{equation}
\begin{equation}\label{op:quintuplet}
\mathcal{O}_\Omega = 
(\overline{\tilde L_{L}} \tau^a L_{L})M^\Omega_{c,ab} (\tilde \phi^\dagger \tau^b \phi) \Omega^{c} ,
\quad {\rm with} \;\; a, b = 1,0,-1, \;\; c =2, 1, 0, -1, -2 , 
\end{equation}
where a sum on repeated indices is understood and we have omitted family indices. 
$M^{\rm H}$ are matrices with only non-zero entries for $a+b=0$ if ${\rm H}=\Delta$ or $\chi$, 
and for $a+b+c=0$ when ${\rm H}=\Sigma$ or $\Omega$: 
\begin{equation}\label{matrix:doublettriplet}
M^\Delta_{ab} = \begin{pmatrix} 0 & 0 & 1\\ 0 & 1 & 0\\ 1 & 0 & 0 \end{pmatrix}; \qquad 
M^\chi_{ab} = \begin{pmatrix} 0 & 0 & 1\\ 0 & -1 & 0\\ 1 & 0 & 0 \end{pmatrix}; \nonumber
\end{equation}
\begin{equation}\label{matrix:quadruplet}
M^\Sigma_{\frac{3}{2},ab} = \begin{pmatrix} 0& 0\\ 0 & 0\\ 0 & -1\end{pmatrix}, 
\qquad 
M^\Sigma_{\frac{1}{2},ab} = \begin{pmatrix} 0& 0\\ 0 & -\sqrt{\frac{2}{3}}\\ \frac{1}{\sqrt{3}} & 0\end{pmatrix}, 
\qquad 
M^\Sigma_{-c,-a -b} = - M^\Sigma_{c,a b} \; ;  \nonumber
\end{equation}
\begin{equation}\label{matrix:quintuplet}
M^\Omega_{2,ab} = \begin{pmatrix} 0 & 0 & 0\\ 0 & 0 & 0\\ 0 & 0 & 1 \end{pmatrix}, 
\qquad 
M^\Omega_{1,ab} = \begin{pmatrix} 0 & 0 & 0\\ 0 & 0 & \frac{1}{\sqrt{2}}\\ 0 & \frac{1}{\sqrt{2}} & 0 \end{pmatrix},
\qquad 
M^\Omega_{0,ab}=\begin{pmatrix} 0 & 0 & \frac{1}{\sqrt{6}}\\ 0 & \sqrt{\frac{2}{3}} & 0\\ \frac{1}{\sqrt{6}} & 0 & 0 \end{pmatrix},
\nonumber 
\end{equation}
\begin{equation}\label{matrix:quintupletbis}
M^\Omega_{-c,-a -b} = M^\Omega_{c,a b} \; .
\end{equation}
After EWSB the resulting Yukawa interactions in Figure \ref{table:YukawaRules} write 
\renewcommand{\arraystretch}{2.0}
\begin{table}[t]
\begin{center}
{\small
\begin{tabular}{ l c c c c c } 
\cbottomrule
\crowcolor ${\mathcal Vertex}$ & Singlet & Doublet & Triplet & Quadruplet & Quintuplet \\
\ctoprule
$\alpha_{ij}^L$ & 0 & $-\dfrac{v}{\Lambda} c_{\chi ij}^{(2)}$ & $-c_{\Delta ij}$ & $\dfrac{v}{\Lambda}c_{\Sigma ij}$ 
& $\dfrac{v^2}{\Lambda^2}c_{\Omega ij}$ \\
\cbottomrule
\crowcolor $\alpha_{ij}^R$ & $c_{\kappa ij}$ & $\dfrac{v}{\Lambda}c_{\chi ij}^{(1)}$ & 0 & 0 & 0 \\
\ctoprule
$\beta_{ij}$ & 0 & $\dfrac{v}{\Lambda}\dfrac{c_{\chi ij}^{(2)}}{\sqrt{2}}$ & $\dfrac{c_{\Delta ij}}{\sqrt{2}}$ 
& $-\dfrac{v}{\Lambda}\dfrac{c_{\Sigma ij}}{\sqrt{3}}$ & $-\dfrac{v^2}{\Lambda^2}\dfrac{c_{\Omega ij}}{2}$ \\
\cbottomrule
\end{tabular}}
\caption{Trilinear Scalar-Fermion-Fermion ($\cal SFF$) couplings for the different multiplet assignments in Eq. (\ref{eq:yuk}).}
\label{table:sff}
\end{center}
\end{table}
\begin{equation}
\label{eq:yuk}
\frac{c_{{\rm H} ij}}{\Lambda^{n_{\rm H}}} \mathcal{O}_{{\rm H} ij} \rightarrow \left(\alpha_{ij}^L \overline{{l}^{c}_{L i}} l_{L j} + 
\alpha_{ij}^R \overline{{l}^{c}_{R i}} l_{R j}\right) {\rm H}^{++} + 
\beta_{ij} \left( \overline{{\nu}^{c}_{L i}} l_{L j} + \overline{{l}^{c}_{L i}} \nu_{L j} \right) {\rm H}^{+} + \cdots ,
\end{equation}
\noindent
where the couplings $\alpha$ and $\beta$ are in general symmetric, flavor-dependent 
and suppressed by powers of $v/\Lambda$, as shown in Table \ref{table:sff}. 
Thus, doubly-charged scalars can 
always couple to same-sign charged lepton pairs in a gauge invariant way independently 
of the EW multiplet they belong to, although in general with suppressed coefficients. 
On the other hand, doubly and singly-charged scalar 
decays are a priori related, even though in practice 
these relations only have phenomenological implications in 
quite specific models, as we shall argue later. 

There can be also operators of the same order but, for instance, quadratic in the scalar fields. 
However, they are in general further suppressed. For example, 
in the quadruplet case the $LL$ interaction in Eq. (\ref{eq:yuk}) can be also 
obtained from the dimension-5 operator 
$\mathcal{O}_{\Sigma\otimes\Sigma} = (\Sigma^\dagger O_a\Sigma)(\overline{\tilde{L}}_L\tau^a L_L)$ 
(where $O_a$ are $4 \times 4$ matrices projecting the $\Sigma\otimes\Sigma$ 
product into the triplet representation), once the neutral $\Sigma$ component 
gets a VEV, $\langle \Sigma^0 \rangle = v_\Sigma$. 
However, this VEV has to be rather small 
($v_\Sigma<$ few GeV) 
in order to satisfy, for instance, the constraint on the rho parameter 
($\rho = 1.0004^{+0.0003}_{-0.0004}$ at the 95 \% C.L. \cite{Beringer:1900zz})
\footnote{\label{rho} 
As can be derived from its generic expression (to lowest order in perturbation theory) 
\begin{equation}
\rho = 
\frac{\sum_k \left[T_k(T_k+1)-Y_k^2\right]v_k^2}{\sum_k 2Y_k^2v_k^2} , \nonumber
\end{equation}
where $k$ labels the scalar multiplets in the model, 
and $T_k$, $Y_k$ and $v_k$ are the corresponding isospin, 
hypercharge and VEV, respectively.}. 
What in general justifies neglecting the contribution of this operator.

%
%
\vspace{-0.1cm}
\section{Gauge scalar interactions}
\label{sec:gauge}

\begin{figure}[]
\begin{center}
\begin{tabular}{c}
\includegraphics[width=0.18\columnwidth]{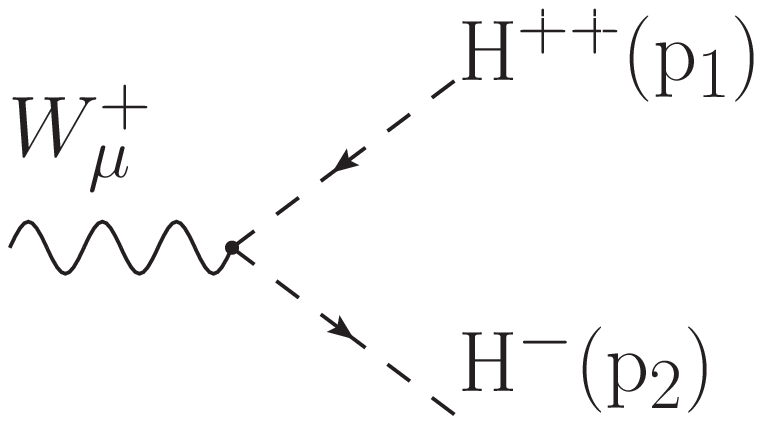} 
\raisebox{0.7cm}{$ic_{W}(p_\mu^1-p_\mu^2)$}
\end{tabular}
\begin{tabular}{cc}
\includegraphics[width=0.22\columnwidth]{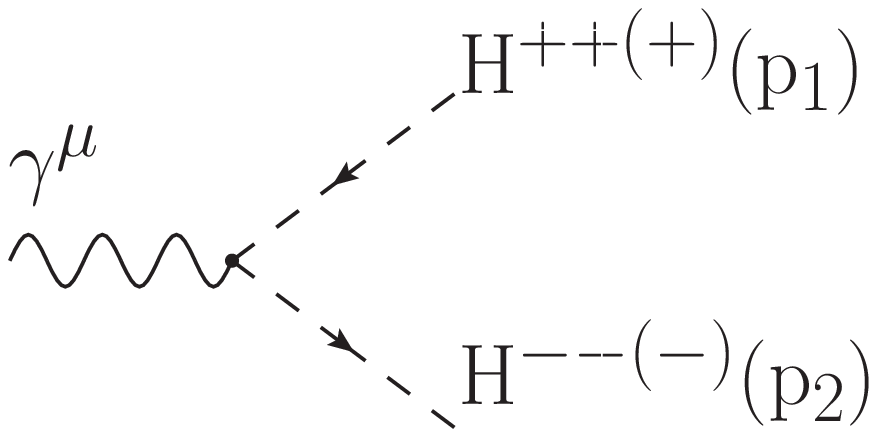} 
\raisebox{0.7cm}{$ic^{(\prime)}_{\gamma}(p_\mu^1-p_\mu^2)$} 
& \includegraphics[width=0.22\columnwidth]{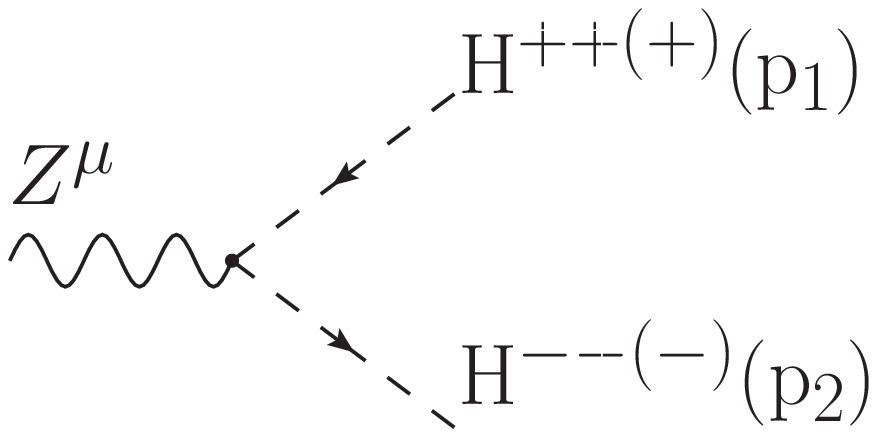} \raisebox{0.7cm}{$ic^{(\prime)}_{Z}(p_\mu^1-p_\mu^2)$}
\end{tabular}
\caption{Feynman diagrams and rules for gauge trilinear interactions of doubly ($c$) and singly ($c'$) charged scalars. 
The arrows indicate the LN flow; whereas the H$^{++(+)}$ ($p_1$) and H$^{--(-)}$ ($p_2$) momenta 
are leaving the vertex.}
\label{fig:ThreeFieldsRules}
\end{center}
\end{figure}

\begin{figure}[]
\begin{center}
\begin{tabular}{ccc}
\hspace{-0.29cm}\includegraphics[width=0.225\columnwidth]{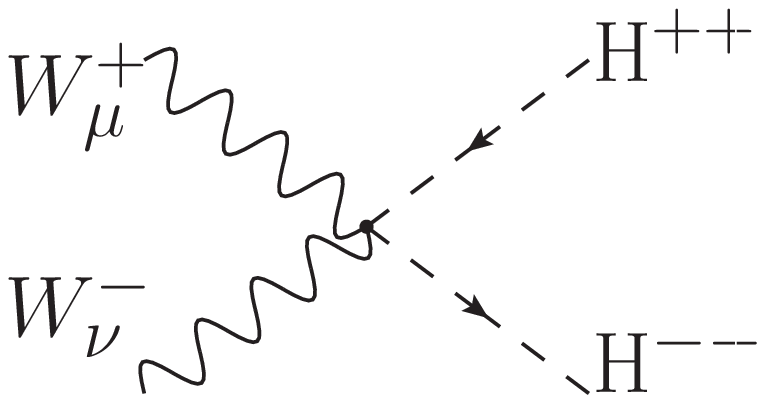} 
\raisebox{0.7cm}{$ic_{WW}g_{\mu\nu}$} & 
\hspace{-0.29cm}\includegraphics[width=0.225\columnwidth]{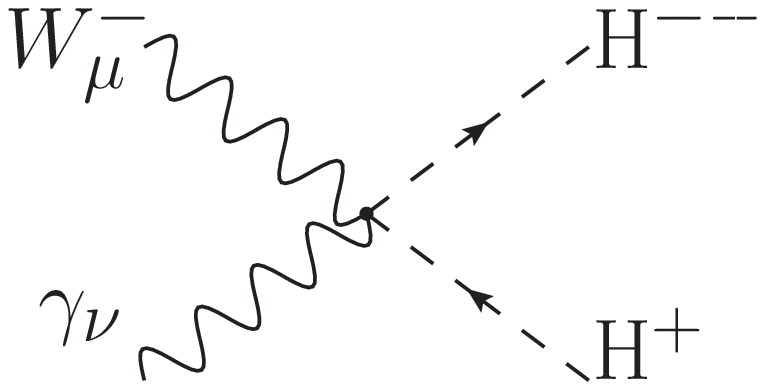} 
\raisebox{0.7cm}{$ic_{W\gamma}g_{\mu\nu}$} 
& \hspace{-0.31cm}\includegraphics[width=0.225\columnwidth]{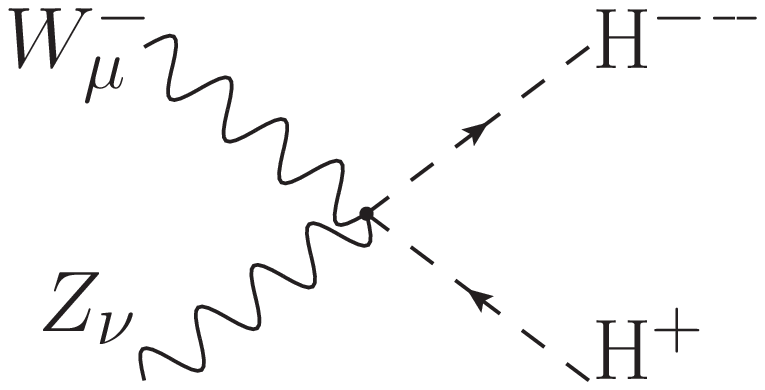} \raisebox{0.7cm}{$ic_{WZ}g_{\mu\nu}$} \\ 
\includegraphics[width=0.2\columnwidth]{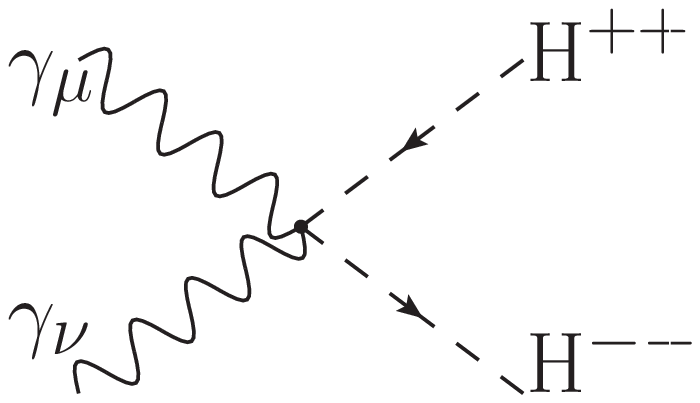} 
\raisebox{0.7cm}{$2ic_{\gamma\gamma}g_{\mu\nu}$} 
& \includegraphics[width=0.2\columnwidth]{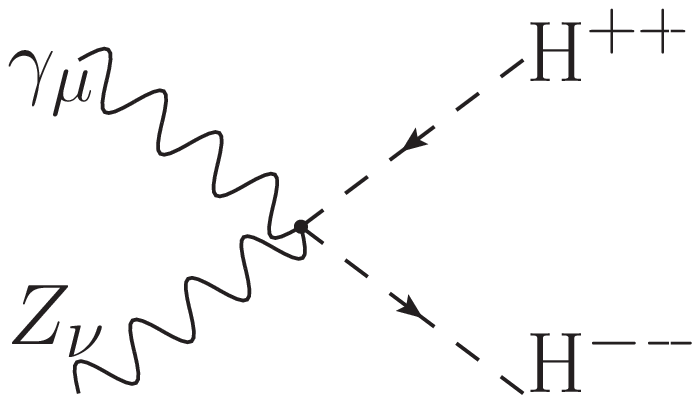} 
\raisebox{0.7cm}{$ic_{\gamma Z}g_{\mu\nu}$} &
\includegraphics[width=0.21\columnwidth]{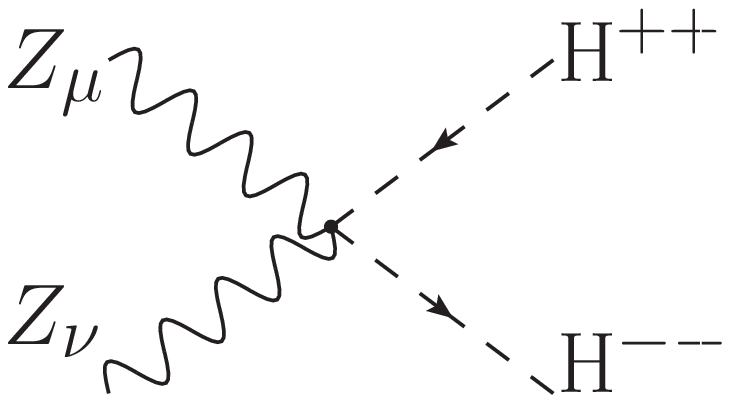} \raisebox{0.7cm}{$2ic_{ZZ}g_{\mu\nu}$} 
\end{tabular}
\caption{Feynman diagrams and rules for gauge quartic interactions of doubly and singly-charged scalars. 
The arrows indicate the LN flow.}
\label{fig:FourFieldsRules}
\end{center}
\end{figure}

Scalar multiplets, ${\rm H}$, with doubly-charged components, ${\rm H}^{++}$, transform 
non-trivially under the EW gauge group and thus couple to 
$\gamma , Z$ and $W$ (except in the singlet case which only 
has neutral interactions). The explicit form of the gauge couplings 
is derived from the corresponding kinetic Lagrangian 
\begin{equation}
\mathcal{L}^K = \left(D^\mu{\rm H}\right)^\dagger D_\mu{\rm H} \; ,
\end{equation}
where the action of the covariant derivative $D_\mu$ reads (in standard notation)
\begin{align}
D_\mu{\rm H}& = \left(\partial_\mu+ig{\vec{T}}\cdot{\vec{W}_\mu}+ig'YB_\mu\right){\rm H} \nonumber \\
& = \left(\partial_\mu + \frac{ig}{\sqrt{2}}\left(T^+W_\mu^++T^- W_\mu^-\right) 
+ \frac{ig}{c_W}\left(T_3-s_W^2Q\right)Z_\mu + ieQA_\mu\right){\rm H} \; ,
\end{align}
with $s_W$ ($c_W$) the sine (cosine) of the EW mixing angle, $s_W = g' /\sqrt{g^2 +g'^2}$, 
$e = g s_W$ the (positive) electromagnetic gauge coupling and $Q = T_3 + Y$ the electric charge operator. 
In particular, expanding $\mathcal{L}^K$ and reordering terms, the trilinear and quartic 
gauge couplings involved in the calculation of the pair and associated production of 
doubly-charged scalars can be written 
\begin{align}
\mathcal{L}^K & \rightarrow 
\left\{ i \frac{g}{\sqrt 2} \sqrt{(T-Y+2)(T+Y-1)} W^{-}_\mu \left[ {\rm H}^{++}(\partial ^\mu {\rm H}^{-}) 
- (\partial ^\mu {\rm H}^{++}){\rm H}^{-} \right] \right. \nonumber \\
& + i \left[ 2 e A_\mu + \frac{g}{c_W} (2-Y-2s_W^2) Z_\mu  \right]
{\rm H}^{++}(\partial ^\mu {\rm H}^{--}) \nonumber \\
& + \left. i \left[ e A_\mu + \frac{g}{c_W} (1-Y-s_W^2) Z_\mu  \right]
{\rm H}^{+}(\partial ^\mu {\rm H}^{-}) + {\rm h.c.} \right\} \nonumber \\ 
& + g^2 \left[ T(T+1) -(2-Y)^2 \right] 
W^{+}_\mu W^{- \mu} {\rm H}^{++}{\rm H}^{--} \nonumber \\ 
& + \left\{ \frac{g}{\sqrt 2} \sqrt{(T-Y+2)(T+Y-1)} W^{-}_\mu 
\left[ 3 e A^\mu + \frac{g}{c_W} (3-2Y-3s_W^2) Z^\mu \right] {\rm H}^{++}{\rm H}^- + {\rm h.c.} \right\} \nonumber \\ 
& + \left[ 2 e A_\mu + \frac{g}{c_W} (2-Y-2s_W^2) Z_\mu  \right] 
\left[ 2 e A^\mu + \frac{g}{c_W} (2-Y-2s_W^2) Z^\mu  \right] {\rm H}^{++}{\rm H}^{--} \, .  
\label{lagrangian}
\end{align}
The first two lines describe the $s-$channel exchange of gauge bosons \cite{delAguila:2013yaa}; 
whereas all of them enter in the calculation of the vector-boson fusion (VBF) contribution 
(see next section). These couplings depend on the type of multiplet, {\it i.e.}, on $T$ and $Y$, 
the doubly-charged scalar belongs to, as do the corresponding cross-sections. 
In Eq. (\ref{lagrangian}) we have used Eq. (\ref{chargerelation}) but omitting superindices 
for easy reading. The doubly (2) and singly (1) charges have been also made explicit. 
In Figures (Tables) \ref{fig:ThreeFieldsRules} and \ref{fig:FourFieldsRules} 
we gather the Feynman diagrams and rules (couplings) 
for the scalar multiplets satisfying Eq. (\ref{chargelessequaltwo}), which are discussed below 
for illustration. 
\begin{table}[]
\begin{center}
{\small
\begin{tabular}{ l c c c c c } 
\cbottomrule
\crowcolor ${\mathcal Vertex}$ & Singlet & Doublet & Triplet & Quadruplet & Quintuplet \\
\ctoprule
$c_{W}$ & 0 & $\frac{g}{\sqrt{2}}$ & $g$ & $\sqrt{\frac{3}{2}}g$ & $\sqrt{2}g$ \\
\cbottomrule
\crowcolor $c_{\gamma}$ & $2e$ & $2e$ & $2e$ & $2e$ & $2e$ \\
\ctoprule
$c_{Z}$ & $-2\dfrac{g}{c_W}s_W^2$ & $\dfrac{g}{2c_W}\left(1-4s_W^2\right)$ 
& $\dfrac{g}{c_W}\left(1-2s_W^2\right)$ & $\dfrac{g}{2c_W}\left(3-4s_W^2\right)$ 
& $\dfrac{2g}{c_W}\left(1-s_W^2\right)$ \\
\cbottomrule
\crowcolor $c'_{\gamma}$ & $0$ & $e$ & $e$ & $e$ & $e$ \\
\ctoprule
$c'_{Z}$ & $0$ & $-\dfrac{g}{2c_W}\left(1+2s_W^2\right)$ & $-\dfrac{g}{c_W}s_W^2$ 
& $\dfrac{g}{2c_W}\left(1-2s_W^2\right)$ & $\dfrac{g}{c_W}\left(1-s_W^2\right)$ \\
\cbottomrule
\end{tabular}
}
\caption{Trilinear Scalar-Scalar-Vector ($\cal SSV$) couplings for doubly ($c$) and singly ($c'$) charged scalars.}
\label{table:ssv}
\end{center}
\end{table}
\begin{table}[]
\begin{center}
\begin{adjustwidth}{-0.28cm}{}
{\small
\begin{tabular}{ l c c c c c } 
\cbottomrule
\crowcolor 
${\mathcal Vertex}$ & Singlet & Doublet & Triplet & Quadruplet & Quintuplet \\
\ctoprule
$c_{WW}$ & 0 & $\dfrac{g^2}{2}$ & $g^2$ & $\dfrac{3g^2}{2}$ & $2g^2$ \\
\cbottomrule
\crowcolor 
$c_{W\gamma}$ & 0 & $\dfrac{3}{\sqrt{2}}eg$ & $3eg$ & $3\sqrt{\dfrac{3}{2}}eg$ & $3\sqrt{2}eg$ \\
\ctoprule
$c_{W Z}$ & 0 & $\dfrac{-3g^2 s_W^2}{\sqrt{2}c_W}$ & $\dfrac{g^2}{c_W}\left[1-3s_W^2\right]$ 
& $\sqrt{\dfrac{3}{2}}\dfrac{g^2}{c_W}\left[2-3s_W^2\right]$ & $\dfrac{3\sqrt{2}}{c_W}g^2\left[1-s_W^2\right]$ \\
\cbottomrule
\crowcolor 
$c_{\gamma\gamma}$ & $4e^2$ & $4e^2$ & $4e^2$ & $4e^2$ & $4e^2$ \\
\ctoprule
$c_{\gamma Z}$ & $-8e^2\dfrac{s_W}{c_W}$ & $\dfrac{2eg}{c_W}\left[1-4s_W^2\right]$ 
& $\dfrac{4eg}{c_W}\left[1-2s_W^2\right]$ & $\dfrac{2eg}{c_W}\left[3-4s_W^2\right]$ 
& $\dfrac{8eg}{c_W}\left[1-s_W^2\right]$ \\
\cbottomrule
\crowcolor 
$c_{ZZ}$ & $4g^2\dfrac{s_W^4}{c_W^2}$ & $\dfrac{g^2}{4c_W^2}\left[1-4s_W^2\right]^2$ 
& $\dfrac{g^2}{c_W^2}\left[1-2s_W^2\right]^2$ & $\dfrac{g^2}{4c_W^2}\left[3-4s_W^2\right]^2$ 
& $\frac{4g^2}{c_W^2}\left[1-s_W^2\right]^2$ \\
\ctoprule
\end{tabular}
}
\caption{Quartic Scalar-Scalar-Vector-Vector ($\cal SSVV$) couplings for VBF doubly-charged pair and associated production.}
 \end{adjustwidth}
\label{table:ssvv}
\end{center}
\end{table}
Quartic couplings involving neutral scalars ${\rm H}^0$, {\it i.e.}, for $T^{\rm H} \geq | T_3^{{\rm H}^{++}} - 2 |$, 
also mediate LNV doubly-charged scalar decays once the LN = 2 neutral component gets a VEV, 
\begin{equation}
\label{WWH}
\mathcal{L}^K \rightarrow \frac{g^2}{2} \sqrt{(T+Y)(T+Y-1)(T-Y+2)(T-Y+1)} 
W^{-}_\mu W^{- \mu} {\rm H}^{++} \langle {\rm H}^{0} \rangle \, .
\end{equation}
Multiplets without neutral components can also decay into $W$ pairs by mixing with other multiplets 
with a neutral component developing a VEV 
\footnote{This may be expected in generic ultraviolet completions. 
As a matter of fact, the effective operators in Eq. (\ref{op:quintuplet}) 
and the effective coupling to $W$ pairs can be obtained from 
renormalizable theories with further scalars, in particular with a heavy triplet 
and/or singlet, after integrating them out \cite{delAguila:2011gr,delAguila:2012nu}.}, 
or through quantum corrections. 
In order to establish LNV both types of decays ${\rm H}^{\pm\pm} \rightarrow l^\pm l^\pm , W^\pm W^\pm$ 
must be observed. Otherwise, the scalar LN could be just 2 in the former case or 0 
in the latter one, but still conserved. In general, it makes sense to look for decays into 
lepton as slow as into gauge boson pairs because although the decay into vector 
bosons is proportional to a VEV which turns out to be minuscule, decays into 
same-sign charged lepton pairs are stringently constrained by current limits on 
lepton flavor violation. 

%
%

\section{Doubly-charged scalar production}
\label{sec:prod}

\begin{figure}[t]
\begin{center}
\includegraphics[width=0.49\columnwidth]{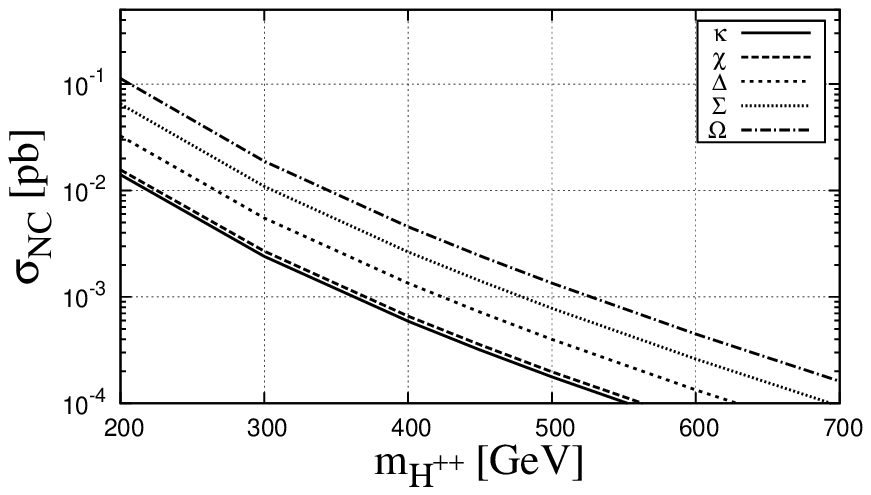}
\includegraphics[width=0.49\columnwidth]{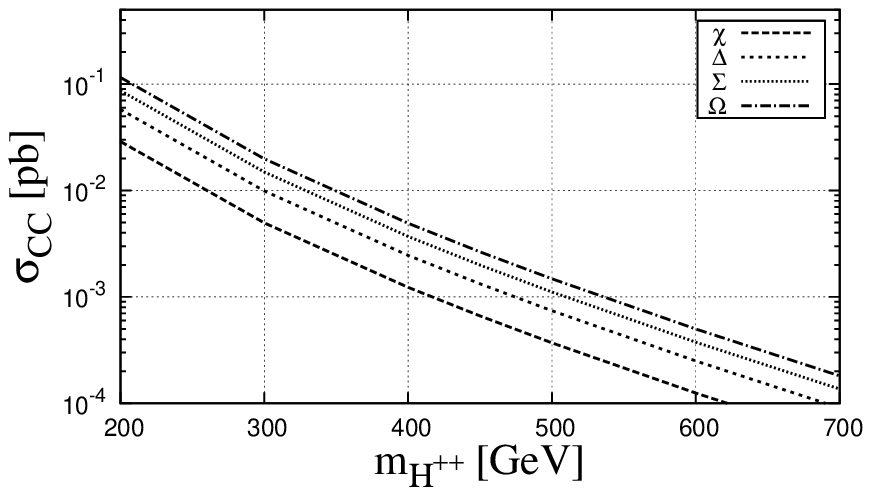}
\end{center}
\caption{Doubly-charged scalar pair (left) and associated 
(right) production at the LHC for $\sqrt{s} = 8$ TeV, with scalars ${\rm H}$ 
belonging to a real quintuplet $\Omega$, a quadruplet $\Sigma$, a triplet $\Delta$, 
a doublet $\chi$ or a singlet $\kappa$ 
with hypercharges 0, 1/2, 1, 3/2 and 2, respectively.}
\label{fig:xsecs}
\end{figure}

Doubly-charged scalars are pair produced with EW strength 
through the $s-$channel exchange of photons and $Z$ bosons, 
$p p \rightarrow \gamma^* / Z^* \rightarrow {\rm H}^{++}{\rm H}^{--}$ 
\footnote{They can be also singly produced through the effective 
coupling  ${\rm H}^{\pm\pm}W^{\mp}_\mu W^{\mu \mp}$ in Eq. (\ref{WWH}). 
Although, due to the stringent constraints on its size, for instance, implied 
by the measured value of the $\rho$ parameter (see footnote \ref{rho}), this 
production mechanism is in general suppressed to a negligible level, 
unless a bizarre cancellation is invoked to avoid these bounds \cite{Chiang:2012dk}.}. 
Similarly, its associated production with a singly-charged 
scalar proceeds through $W$ exchange, 
$p p \rightarrow W^{\pm *} \rightarrow {\rm H}^{\pm\pm}{\rm H}^\pm$.
Both cross-sections depend on the quantum numbers of the 
scalar multiplet the doubly-charged scalar belongs to, as 
do the corresponding couplings in Eq. (\ref{lagrangian}). 
In Figure \ref{fig:xsecs} 
we plot them as a function of the doubly-charged scalar mass 
$m_{{\rm H}^{++}}$ for the five cases in Figure \ref{fig:ThreeFieldsRules} 
and Table \ref{table:ssv} and for $\sqrt s = 8$ TeV 
(the corresponding cross-sections for $\sqrt s = 14$ TeV are shown 
in \cite{delAguila:2013yaa} 
\footnote{The scalar triplet cross-sections at 7 TeV 
are plotted, for example, in \cite{Chatrchyan:2012ya,ATLAS:2012hi}. 
For an earlier comparison of the Tevatron and LHC potential 
see \cite{Akeroyd:2005gt}.}). 

\begin{figure}[]
\begin{center}
\includegraphics[width=0.25\columnwidth]{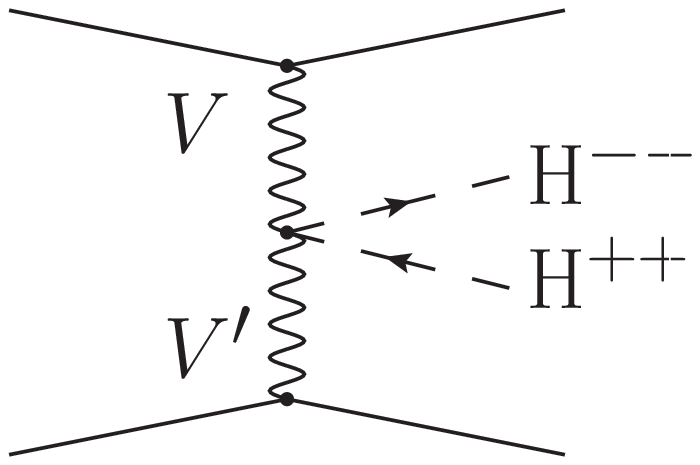}\hspace{1cm}
\includegraphics[width=0.25\columnwidth]{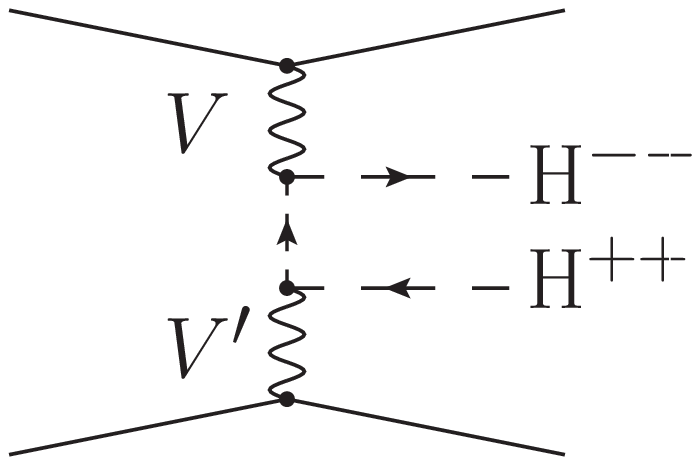}\hspace{1cm}
\includegraphics[width=0.25\columnwidth]{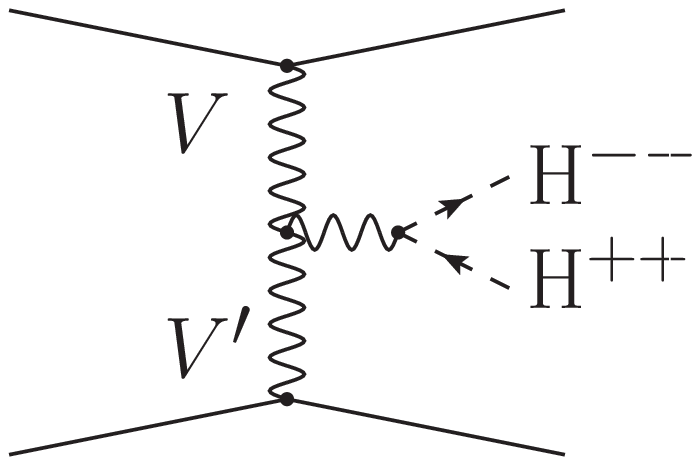}\hspace{1cm}
\end{center}
\caption{Feynman diagrams contributing to VBF doubly-charged scalar pair production.}
\label{fig:vbfdiagrams}
\end{figure}

Both final states can be also produced through VBF but accompanied by two extra jets,  
$pp \rightarrow {\rm H}^{++} {\rm H}^{--}jj$, ${\rm H}^{\pm\pm} {\rm H}^{\mp}jj$. 
These processes are sub-leading as expected from gauge-coupling power counting. 
The contributing diagrams are depicted in Figure \ref{fig:vbfdiagrams}. 
Although this mechanism is enhanced because the initial partons are both valence quarks, 
its size stays below 10 \% of the $s-$channel production, being almost negligible for 
low scalar masses. 
\begin{figure}[]
\begin{center}
\includegraphics[width=0.49\columnwidth]{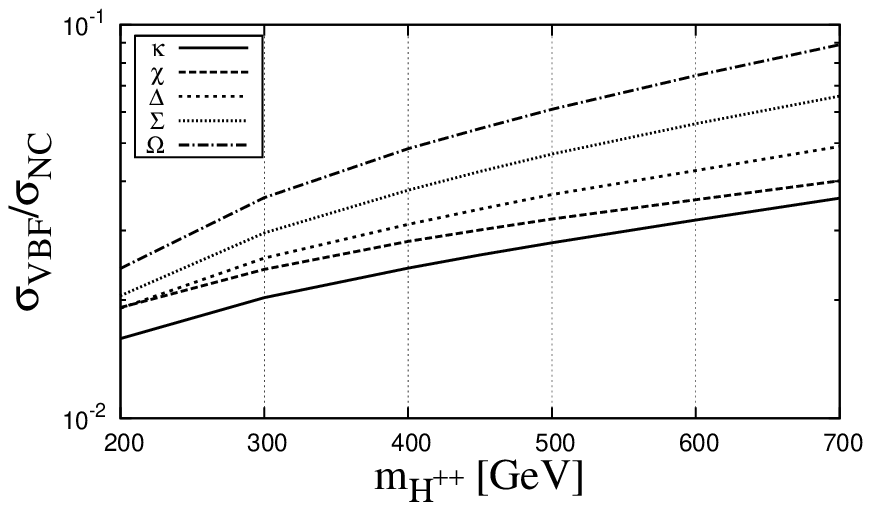}
\includegraphics[width=0.49\columnwidth]{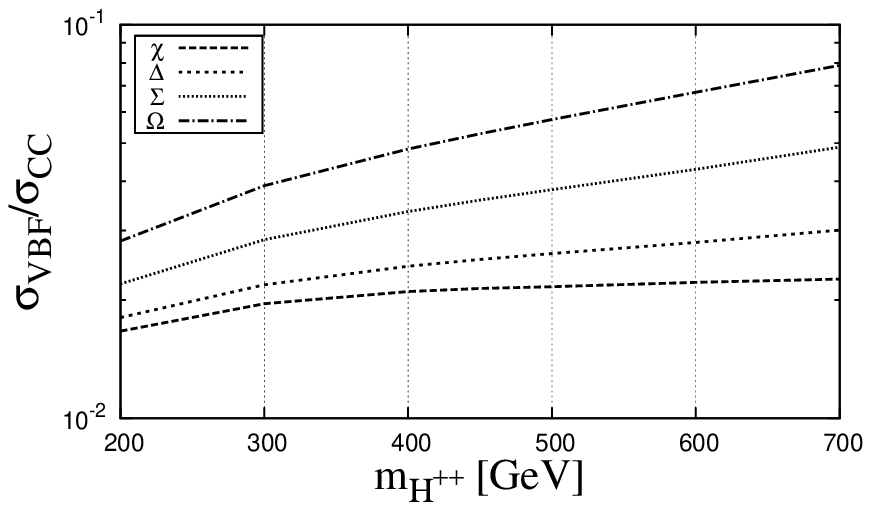}
\end{center}
\caption{Ratio of the VBF to the pair (left) and associated (right) production 
cross-sections as a function of the doubly-charged scalar mass for the same multiplets and 
energy as in Figure \ref{fig:xsecs}.}
\label{fig:vbf}
\end{figure}
In Figure \ref{fig:vbf} we plot the ratio of the VBF to the $s-$channel production cross-section 
for the same scalar multiplets as in Figure \ref{fig:xsecs} 
\footnote{The ratio for pair production but including only the VBF of two photons, 
$\gamma\gamma \rightarrow {\rm H}^{++} {\rm H}^{--}$, is quite similar as previously shown 
in \cite{Han:2007bk}. 
These partonic cross-sections diverge when the 
photon is emitted collinearly and then are sensitive to the limit on the corresponding 
partonic transverse momentum. We assume as a conservative value $p_T^j > 10$ GeV 
throughout the paper.}. 
As can be observed in the figure VBF starts to be important only for large masses, 
when the valence quark parton distribution functions (PDFs) are relatively larger. 
Anyway, this production mechanism is always present and should be taken into account, 
although it is possible to separate the corresponding events by requiring two forward 
extra jets. (Collinear $\gamma$ production can be calculated using the 
Weizsaecker-Williams approximation 
\cite{Williams:1934ad,Drees:1994zx}, giving also similar contributions \cite{Han:2007bk}.) 

Besides, there can be further NP contributions, although in general further suppressed. 
For example, effective operators contributing to these processes  are suppressed by at 
least two powers of the cutoff scale $\Lambda$ (see Appendix \ref{effectiveproduction}). 
As a matter of fact, these contact interactions arise naturally in the context, for instance, of 
non-minimal Composite Higgs Models \cite{Gripaios:2009pe}. 
On the other hand, possibly $s-$channel contributions are in addition suppressed by 
small far off-shell propagators 
\footnote{This is also the case for the Higgs $s$-channel exchange, whose rate is much 
smaller than the one from the exchange of gauge bosons, unless the effective Higgs coupling 
to doubly-charged scalars is unnaturally large 
(equal to $\lambda v$ with $\lambda$ much larger than one).}; 
while $t-$channel contributions are forbidden, since 
doubly-charged scalars do not have trilinear couplings to a quark pair. 

In summary, although there can be a variety of production mechanisms, in general 
the main production cross-sections are fixed by the scalar multiplet quantum numbers 
in Eq. (\ref{lagrangian}); and thus their measurement would allow to determine 
the total isospin and hypercharge of the scalar multiplet which the doubly-charged scalar 
belongs to 
\cite{delAguila:2013yaa,delAguila:2013hla}. 
In the signal simulations below we add the $s-$channel as well as the VBF production, 
multiplying the former by a K-factor equal to 1.25 \cite{Muhlleitner:2003me}. 

%
%

\section{Monte Carlo implementation}
\label{sec:Simulation}

In order to extend the searches of doubly-charged scalars to 
generic models and to perform phenomenological studies 
allowing for LNV signals, we have implemented the 
couplings in Eqs. (\ref{eq:yuk}), (\ref{lagrangian}) and (\ref{WWH}) 
in \texttt{MadGraph5} \cite{Alwall:2011uj} 
\footnote{Further details for its use are given in Appendix \ref{sec:SimulationAndAnalyses}.}.
The explicit expressions of these couplings for the five scalar 
multiplets of lowest isospin and hypercharge containing a doubly-charged component, 
$(T,Y) = (0,2), (1/2,3/2), (1,1), (3/2,1/2), (2,0)$, have been 
included in an \texttt{UFO} model by means of \texttt{FeynRules v1.6} \cite{Christensen:2008py}. 
It can be downloaded from \href{http://cafpe.ugr.es/index.php/pages/other/software}
{\textsf{http://cafpe.ugr.es/index.php/pages/other/software}}. A set of \texttt{Param Cards} for 
\texttt{MadGraph5} can be also found there for all scalar masses 
considered in the simulations in the text. 
The K-factor for the leading pair and associated $s-$channel production 
can be as large as 20-30 \% \cite{Muhlleitner:2003me}. 
This and the VBF contributions can be added at will, 
although the latter are only sizable for large doubly-charged scalar masses. 

In order to take into account the scalar decays mediated by the 
$\mathcal{SFF}$ and $\mathcal{SVV}$ interactions in 
Eqs. (\ref{eq:yuk}) and (\ref{WWH}), respectively, 
it is enough to implement the triplet decay. This is so because 
all those decays can be mimicked by the triplet one, 
being only required an additional rescaling of the corresponding rate. 
In particular, as chirality does not play any role in the subsequent analyses 
\footnote{The helicity of the final leptons can not be measured, 
except eventually for the tau lepton \cite{Sugiyama:2012yw}.}, 
we only need to consider $LL$ interactions, 
{\it i.e.}, $\Delta^{++} \overline{l^c_{L i}} l'_{L j}$ and $\Delta^{+} \overline{l^c_{L i}} \nu'_{L j}$, 
with the six possible lepton combinations  
$ee, e\mu, e\tau, \mu\mu, \mu\tau$ and $\tau\tau$ 
in the doubly-charged case and only three 
$e\nu_e, \mu\nu_\mu$ and $\tau\nu_\tau$ in the singly-charged one, 
for neutrinos manifest as missing energy in the detector and hence 
cannot be distinguished. 
A new parameter \texttt{myyuk} has been introduced to account for the 
Yukawa couplings. The scalar decays into $W$ pairs mediated by 
the interactions $\Delta^{\pm\pm} W^\pm W^\pm$ and $\Delta^\pm W^\pm Z^0$ 
also require the introduction of a new parameter, which we name \texttt{myvev}, 
proportional to the corresponding (LNV) VEV.
These two parameters have been fixed to values which make 
the branching ratios into leptons and into bosons of similar magnitude. 
This guarantees that the total cross-section 
never vanishes, avoiding numerical problems. These values 
also imply a narrow scalar width, and hence the width measured 
from the invariant-mass distribution is dominated by reconstruction effects. 
Allowing for scalar decays into lepton and boson pairs, non-trivial limits 
on LNV, not considered yet in experimental searches, can be obtained.  
We will focus on pair and associated production with one scalar decaying 
into lepton and the other into boson pairs 
\footnote{There is a different and much more involved way to look for LNV: 
performing the usual four lepton analyses and also to search for four vector 
bosons compatible with the same resonance production. This second final 
state, however, can be only disentangled from the background for very low masses 
\cite{Kanemura:2013vxa}.}. 
Although no analysis has been designed to look for these specific LNV processes, 
current searches for doubly-charged resonances decaying only into same-sign charged 
lepton pairs are already sensitive to this channel. 
Given that no event excess has been observed, they can be also used to set 
the first bounds on LNV scalars. 
We discuss them in detail later. 

\texttt{MadGraph5} parton level events 
are passed through \texttt{Pythia v6} 
\cite{Sjostrand:2006za} to include initial and final state radiation, as well as fragmentation and 
hadronization, 
and through \texttt{Delphes v3} \cite{Ovyn:2009tx} for fast-detector simulation. 
Jets are reconstructed using an anti-kt algorithm with \texttt{FastJet v3} \cite{Cacciari:2011ma}. 
Finally, \texttt{MadAnalysis v5} \cite{Conte:2012fm} is used to perform the analyses. 
The full sequence of software for Monte Carlo simulation, which has been 
extensively tested, is available to generalize our results.  

%
%

\section{Current analyses}
\label{sec:analyses}

CMS \cite{Chatrchyan:2012ya} and ATLAS \cite{ATLAS:2012hi} 
have provided limits on doubly-charged scalars decaying into same-sign $e$ and 
$\mu$ pairs using samples with four and three isolated charged leptons, as no 
event excess has been observed. 
In this section we reproduce their results at $\sqrt s =$ 7 TeV using Monte Carlo 
simulations in order to test our codes. 
We mimic the detailed analysis by CMS in \cite{Chatrchyan:2012ya}, 
which besides includes doubly-charged scalar decays into $\tau$ leptons. 
We then estimate the expected bounds for 8 TeV and an integrated luminosity 
of 20 fb$^{-1}$ also assuming that no event excess is observed. 
In this analysis we apply the same cuts and efficiencies as for 7 TeV, although 
the LHC collaborations will certainly optimize both and will provide better limits 
based on real data. However, no large differences should be expected. 
In the next section and as another application, we extend these analyses 
to obtain the corresponding limits on the LNV processes 
$pp\rightarrow {\rm H}^{\pm\pm}{\rm H}^{\mp\mp} \rightarrow {\ell}^\pm {\ell}^\pm W^\mp W^\mp$ and 
$pp\rightarrow {\rm H}^{\pm\pm}{\rm H}^{\mp} \rightarrow {\ell}^\pm {\ell}^\pm W^\mp Z$, 
also estimating for both processes the bounds which shall be eventually 
obtained by the LHC experiments after the next run at 14 TeV. 

In order to compare with data, the SM backgrounds must be also included. 
Since the signal efficiencies for $\sqrt s =$ 7 and 8 TeV are similar 
(we find differences of at most $\sim 10 \%$), 
we assume that this is also the case for the backgrounds and 
estimate them at 8 TeV scaling the CMS 
values in Table 5 in \cite{Chatrchyan:2012ya} by a factor of 
\begin{equation}
\frac{\sigma_8}{\sigma_7} \times \frac{\mathcal{L}_8}{\mathcal{L}_7} \approx 
1.2 \times 4.08 \; ,  
\label{scaling}
\end{equation}
where the first figure is the average of the ratios of the corresponding 
cross-sections for the largest backgrounds in Table \ref{table:CrossSections}, 
Drell-Yan, $W^+ W^-$, $W^\pm Z$, $Z Z$ and $t\bar{t}$ production, 
and the second one is the luminosity ratio 20/4.9. 
\begin{table}[]
\begin{center}
{
\begin{tabular}{ l r r r } 
\ctoprule
${\mathcal Process}$ &  $\sigma$[7 TeV] @ NLO (pb) & $\sigma$[8 TeV] @ NLO (pb) 
& $\sigma$[14 TeV] @ NLO (pb) \\
\cbottomrule
\crowcolor  Drell-Yan & $(21\pm 1)\times 10^2$ & $(25\pm 2)\times 10^2$ & $(48\pm 4)\times 10^2$ \\
\ctoprule
$W^+ W^-$ & $41\pm1$ & $50\pm 2$ & $107\pm 4$ \\
\cbottomrule
\crowcolor $W^\pm Z$ & $17\pm 1$ & $21\pm 1$ & $47\pm 2$ \\ 
\ctoprule
$Z Z$ & $5.5\pm 0.2$ & $6.6\pm 0.2$ & $14.5\pm 0.4$ \\ 
\cbottomrule
\crowcolor $t\bar{t}$ & $123\pm 15$ & $176\pm 22$ & $475\pm 9$ \\ 
\ctoprule
\end{tabular}
}
\caption{\label{table:CrossSections}
Cross-sections for the main backgrounds considered in the analyses, computed at the NLO in QCD. At the parton level, events have been generated using \texttt{aMC@NLO} with the cut $p_T^j>10$ GeV (in addition and only for Drell-Yan, $l^+ l^-$, we require $p_T^l > 20$ GeV, $m_{l^+l^-}>30$ GeV and $\Delta R_{l^+l^-} > 0.4 $). The 5 flavor scheme has been used, and the partonic events linked to Pythia by means of the MC@NLO method \cite{Frixione:2002ik}, with the subsequent decay of the $t\bar{t}$ and di-boson final states into their different decay products.}
\end{center}
\end{table}
The number of observed events is assumed to be equal to the number of 
expected background events. 
(We assume the same at 7 TeV for the scalar masses not gathered in Table 
5 in \cite{Chatrchyan:2012ya}, 
taking also in this case the number of expected 
background events to be equal to the number of events predicted by the SM 
\footnote{We assume that the number of events for $m_{{\rm H}^{\pm\pm}} =$ 500 GeV 
is the same as for 450 GeV in \cite{Chatrchyan:2012ya}, and no background events 
are expected and none is observed after the corresponding selection cuts 
for $m_{{\rm H}^{\pm\pm}} =$ 600 and 700 GeV.}.) 
For the LHC run at 14 TeV we have instead simulated the complete set of backgrounds 
in Table \ref{table:CrossSections} for an integrated luminosity of 
100 fb$^{-1}$, assuming again that the observed number of events 
coincides with the expected number of background events after cuts.

CMS has performed six different analyses using four and three isolated 
charged lepton samples, $\ell \ell \ell \ell$ and $\ell \ell \ell$, 
with $\ell = e, \mu$ and $\tau_h$ (although at least two of them 
must be same-sign electrons or muons). 
In the first three studies doubly-charged scalars are pair produced and assumed 
to decay 100 \% of the time into $\ell^\pm \ell^\pm$, $\ell^\pm \tau^\pm$ and $\tau^\pm \tau^\pm$ in turn. 
The cuts and efficiencies are optimized for each case, and events are generated 
for different scalar masses. 
In Table \ref{table:EfficienciesOnFourLeptonsAt7TeV} we collect the corresponding 
cuts and our estimates of the cumulative efficiencies cut-by-cut for a low (200 GeV) 
and a relatively large (500 GeV) scalar mass for illustration. 
\begin{table}[!Ht]
\begin{center}
{
\begin{tabular}{ l l r r } 
\ctoprule 
\multicolumn{2}{c}{Cuts} & \multicolumn{2}{c}{Efficiencies} \\ [-0.4cm]
\multicolumn{2}{c}{} & $m_{{\rm H}^{\pm\pm}} = 200$ GeV &  $500$ GeV \\ 
\cbottomrule
\multicolumn{4}{>{\columncolor[rgb]{0.9,0.9,0.9}}c}{$\ell^\pm \ell^\pm \ell^\mp \ell^\mp$} \\ [-0.3cm]
\vspace{-0.3cm}
Basic cuts & $p_T^{\ell_{1(2)}} > 20 (10)$ GeV, $|\eta^{\ell}|<2.5$ & 68 & 72 \\
\vspace{-0.3cm}
Total $p_T$ & $\sum p^{\ell}_T > 0.6 m_{{\rm H}^{\pm\pm}} + 130$ GeV & 99 & 100 \\
\vspace{-0.1cm}
Mass window & $m_{\ell^\pm \ell^\pm} \in [0.9 m_{{\rm H}^{\pm\pm}}, 1.1 m_{{\rm H}^{\pm\pm}}]$ & 92 & 89 \\ 
\hline
\vspace{-0.2cm}
Total & & 62 & 64 \\ 
\cbottomrule 
\multicolumn{4}{>{\columncolor[rgb]{0.9,0.9,0.9}}c}{$\ell^\pm \tau^\pm \ell^\mp \tau^\mp$} \\ [-0.3cm]
\vspace{-0.3cm}
Basic cuts & $p_T^{\ell_{1(2)}} > 20 (10)$ GeV, $|\eta^{\ell}|<2.5$ & 16 & 23 \\
\vspace{-0.3cm}
Total $p_T$ & $\sum p^{\ell}_T > m_{{\rm H}^{\pm\pm}} + 100$ or $> 400$ GeV & 82 & 99 \\
\vspace{-0.3cm}
$Z$ veto & $|m_{\ell^\pm \ell^\pm} -m_Z| > 10$ GeV & 85 & 92 \\
\vspace{-0.1cm}
Mass window & $m_{\ell^\pm \ell^\pm} \in [0.5 m_{{\rm H}^{\pm\pm}}, 1.1 m_{{\rm H}^{\pm\pm}}]$ & 81 & 66 \\\hline
\vspace{-0.2cm}
Total & & 9.0 & 14 \\
\cbottomrule
\multicolumn{4}{>{\columncolor[rgb]{0.9,0.9,0.9}}c}{$\tau^\pm \tau^\pm \tau^\mp \tau^\mp$} \\ [-0.3cm]
\vspace{-0.3cm}
Basic cuts & $p_T^{\ell_{1(2)}} > 20 (10)$ GeV, $|\eta^{\ell}|<2.5$ & 3.0 & 5.0 \\
\vspace{-0.3cm}
Total $p_T$ & $\sum p^{\ell}_T > $ 120 GeV & 99 & 100 \\
\vspace{-0.3cm}
$Z$ veto & $|m_{\ell^\pm \ell^\pm} -m_Z| > 50$ GeV & 82 & 86 \\
\vspace{-0.1cm}
$\Delta\phi$ & $\Delta\phi_{\ell^\pm \ell^\pm} < 2.5$ & 80 & 80 \\\hline
\vspace{-0.2cm}
Total & & 2.0 & 3.5 \\
\cbottomrule
\end{tabular}
}
\caption{\label{table:EfficienciesOnFourLeptonsAt7TeV}
Applied cuts to the four isolated charged lepton sample 
$\ell \ell \ell \ell$, with two $\ell = e$ or $\mu$ and the other two 
$e, \mu$ or $\tau_h$, and efficiency percentage 
for each successive cut for the final states 
$\ell^\pm \ell^\pm \ell^\mp \ell^\mp$, $\ell^\pm \tau^\pm \ell^\mp \tau^\mp$ and 
$\tau^\pm \tau^\pm \tau^\mp \tau^\mp$ and two representative scalar masses. 
The basic transverse momentum cuts are imposed on the two leptons, electrons or muons, 
required by the trigger; whereas the transverse momentum sum is over the four charged leptons, 
as the generic pseudo-rapidity cut. 
In the three analyses no background events are expected and no event is observed 
for an integrated luminosity of 4.9 fb$^{-1}$ at $\sqrt s =$ 7 TeV.
}
\end{center}
\end{table}
As pointed out by CMS, the efficiencies slightly increase 
with the scalar mass. On the other hand, the mass window is the most 
effective cut, implying a large reduction of the background.  
In the $\ell^\pm \tau^\pm \ell^\mp \tau^\mp$ and $\tau^\pm \tau^\pm \tau^\mp \tau^\mp$ 
analyses tau decays into hadrons are also taking into account. 
Hadronic tau leptons ($\tau_h$) are tagged by a pure geometrical 
method in \texttt{Delphes}, becoming a jet a potential $\tau_h$ 
if a generated $\tau$ is found within a fixed distance 
$\Delta R$ of the jet axis. 

\begin{table}[t]
\begin{center}
{
\begin{tabular}{ l l r r } 
\ctoprule 
\multicolumn{2}{c}{Cuts} & \multicolumn{2}{c}{Efficiencies} \\ [-0.3cm]
\multicolumn{2}{c}{} & $m_H = 200$ GeV &  $500$ GeV \\ 
\cbottomrule
\multicolumn{4}{>{\columncolor[rgb]{0.9,0.9,0.9}}c}{$\ell^\pm \ell^\pm \ell^\mp \nu_\ell$} \\ 
\vspace{-0.3cm}
Basic cuts & $p_T^{\ell_{1(2)}} > 20 (10)$ GeV, $|\eta^{\ell}|<2.5$ & 78 & 82 \\
\vspace{-0.3cm}
Total $p_T$ & $\sum p^{\ell}_T > 1.1 m_{{\rm H}^{\pm\pm}} + 60$ GeV & 84 & 87 \\
\vspace{-0.3cm}
$Z$ veto & $|m_{\ell^\pm \ell^\pm} -m_Z| > 80$ GeV & 59 & 90 \\
\vspace{-0.3cm}
$\Delta\phi$ & $\Delta\phi_{\ell^\pm \ell^\pm} < m_{{\rm H}^{\pm\pm}} ({\rm GeV}) / 600 + 1.95$ & 86 & 94 \\
Mass window & $m_{\ell^\pm \ell^\pm} \in [0.9 m_{{\rm H}^{\pm\pm}}, 1.1 m_{{\rm H}^{\pm\pm}}]$ & 94 & 93 \\\hline
\vspace{-0.2cm}
Total & & 31 & 56 \\
\cbottomrule
\vspace{-0.2cm}
Expected background & & 0.99 & 0.14 \\
\vspace{-0.2cm}
Observed events & & 2 & 1 \\
\cbottomrule
\multicolumn{4}{>{\columncolor[rgb]{0.9,0.9,0.9}}c}{$\ell^\pm \tau^\pm \ell^\mp \nu_\tau (\tau^\mp \nu_\ell)$} \\ 
\vspace{-0.3cm}
Basic cuts & $p_T^{\ell_{1(2)}} > 20 (10)$ GeV, $|\eta^{\ell}|<2.5$ & 16 & 20 \\
\vspace{-0.3cm}
Total $p_T$ & $\sum p^{\ell}_T > 0.85 m_{{\rm H}^{\pm\pm}} + 125$ GeV & 38 & 48 \\
\vspace{-0.3cm}
$Z$ veto & $|m_{\ell^\pm \ell^\pm} -m_Z| > 80$ GeV & 85 & 93 \\
\vspace{-0.3cm}
$E_T^\text{miss}$ & $E_T^\text{miss} > 20$ GeV & 98 & 99 \\
\vspace{-0.3cm}
$\Delta\phi$ & $\Delta\phi_{\ell^\pm \ell^\pm} < m_{{\rm H}^{\pm\pm}} ({\rm GeV}) / 200 + 1.15$ & 83 & 100 \\
Mass window & $m_{\ell^\pm \ell^\pm} \in 
[0.5 m_{{\rm H}^{\pm\pm}}, 1.1 m_{{\rm H}^{\pm\pm}}]$ & 91 & 89 \\\hline
\vspace{-0.2cm}
Total & & 3.8 & 7.9 \\
\cbottomrule
\vspace{-0.2cm}
Expected background & & 1.51 & 0.18 \\
\vspace{-0.2cm}
Observed events & & 3 & 1 \\
\cbottomrule
\end{tabular}
}
\end{center}
\end{table}
\begin{table}[]
\begin{center}
 \begin{adjustwidth}{-0.6cm}{}
{
\begin{tabular}{ l l r r } 
\cbottomrule
\multicolumn{4}{>{\columncolor[rgb]{0.9,0.9,0.9}}c}{$\tau^\pm \tau^\pm \tau^\mp \nu_\tau$} \\ 
\vspace{-0.3cm}
Basic cuts & $p_T^{\ell_{1(2)}} > 20 (10)$ GeV, $|\eta^{\ell}|<2.5$ & \qquad \quad \quad 4.2 & \qquad \quad 8.3 \\
\vspace{-0.3cm}
Total $p_T$ & $\sum p^{\ell}_T > m_{{\rm H}^{\pm\pm}} - 10$ or $> 200$ GeV & 55 & 91 \\
\vspace{-0.3cm}
$Z$ veto & $|m_{\ell^\pm \ell^\pm} -m_Z| > 50$ GeV & 80 & 85 \\
\vspace{-0.3cm}
$E_T^\text{miss}$ & $E_T^\text{miss} > 40$ GeV & 86 & 97 \\
\vspace{-0.3cm}
$\Delta\phi$ & $\Delta\phi_{\ell^\pm \ell^\pm} < 2.1$ & 84 & 84 \\
Mass window & $m_{\ell^\pm \ell^\pm} \in 
[0.5 m_{{\rm H}^{\pm\pm}} - 20\ {\rm GeV}, 1.1 m_{{\rm H}^{\pm\pm}}]$ & 76 & 42 \\\hline
\vspace{-0.2cm}
Total & & 1.0 & 2.2 \\
\cbottomrule
\vspace{-0.2cm}
Expected background & & 1.51 & 0.18 \\
\vspace{-0.2cm}
Observed events & & 3 & 1 \\
\cbottomrule
\end{tabular}
}
\caption{\label{table:EfficienciesOnThreeLeptonsAt7TeV}
Applied cuts to the three isolated charged lepton sample 
$\ell \ell \ell$, with two $\ell = e$ or $\mu$ and the third one 
$e, \mu$ or $\tau_h$, and efficiency percentage 
for each successive cut for the final states 
$\ell^\pm \ell^\pm \ell^\mp \nu_\ell$, $\ell^\pm \tau^\pm \ell^\mp \nu_\tau (\tau^\mp \nu_\ell)$ and 
$\tau^\pm \tau^\pm \tau^\mp \nu_\tau$ and two representative scalar masses. 
The basic transverse momentum cuts are imposed on the two leptons, electrons or muons, required by 
the trigger; whereas the transverse momentum sum is over the three charged leptons, 
as the generic pseudo-rapidity cut. 
The expected background events as well as the observed ones 
for an integrated luminosity of 4.9 fb$^{-1}$ at $\sqrt s =$ 7 TeV are also listed.}
\end{adjustwidth}
\end{center}
\end{table}
Analogously, in Table \ref{table:EfficienciesOnThreeLeptonsAt7TeV} 
we gather the corresponding cuts and estimated cumulative efficiencies 
for the three charged lepton sample and doubly-charged 
scalar associated production. 
Similarly to the doubly-charged scalar, the singly-charged scalar 
${\rm H}^\pm$ is assumed to decay 100 \% of the time into $\ell^\pm \nu_\ell$, 
$\ell^\pm \nu_\tau (\tau^\pm \nu_\ell)$ and $\tau^\pm \nu_\tau$ in turn.
In order to be conservative we make use of the $\ell^\pm \tau^\pm \tau^\mp \nu_\ell$ 
efficiency for 
$\ell^\pm \tau^\pm \ell^\mp \nu_\tau$, too, although the former is smaller due to 
the required extra tau branching ratio into electrons and muons.   
As emphasized in Ref. \cite{Chatrchyan:2012ya}, the efficiencies 
for the three-lepton analyses are near a factor 2 smaller than 
for the corresponding four-lepton analyses. 

Using the estimated efficiencies for seven doubly-charged scalar masses, 
$m_{{\rm H}^{\pm\pm}} =$ 200, 300, 400, 450, 500, 600 and 700 GeV, 
\footnote{Efficiencies for intermediate masses can be obtained 
by interpolation.}
and the expected background and observed number of events, 
we can draw the corresponding exclusion plots as no event excess 
has been observed. 
In Figure \ref{fig:7limits} from top to bottom 
we plot the 95 \% C.L. limits (see Appendix \ref{statistics} for the pertinent definitions) 
for the analyses in Tables \ref{table:EfficienciesOnFourLeptonsAt7TeV} 
(left) and \ref{table:EfficienciesOnThreeLeptonsAt7TeV} (right) 
\footnote{Exclusion limits for intermediate masses are obtained by splines interpolation.}. 
\begin{figure}[]
\begin{tabular}{cc}
\includegraphics[width=0.49\columnwidth]{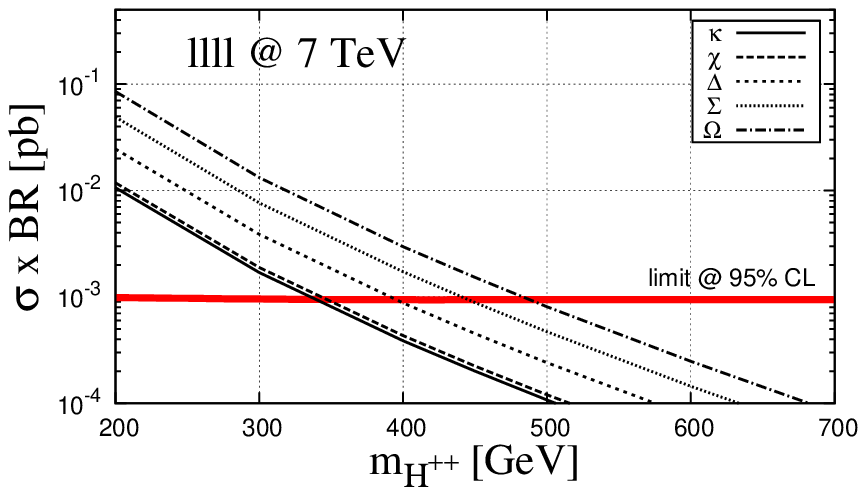} 
& \includegraphics[width=0.49\columnwidth]{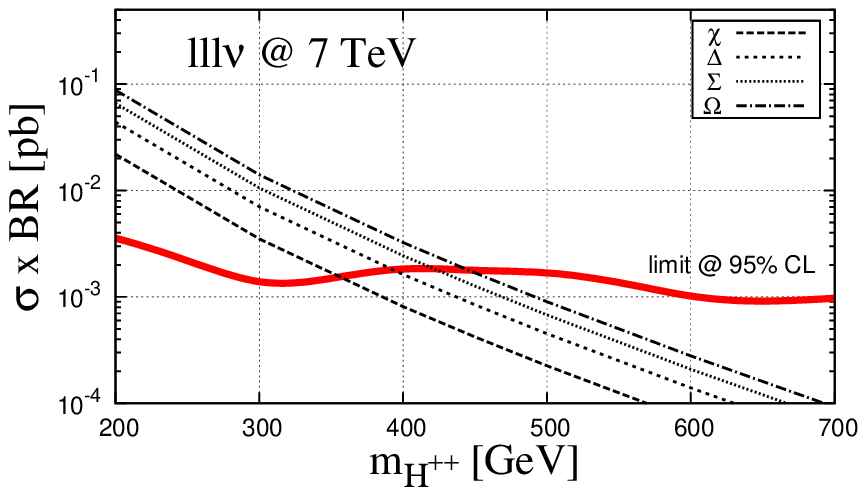} \\ 
\includegraphics[width=0.49\columnwidth]{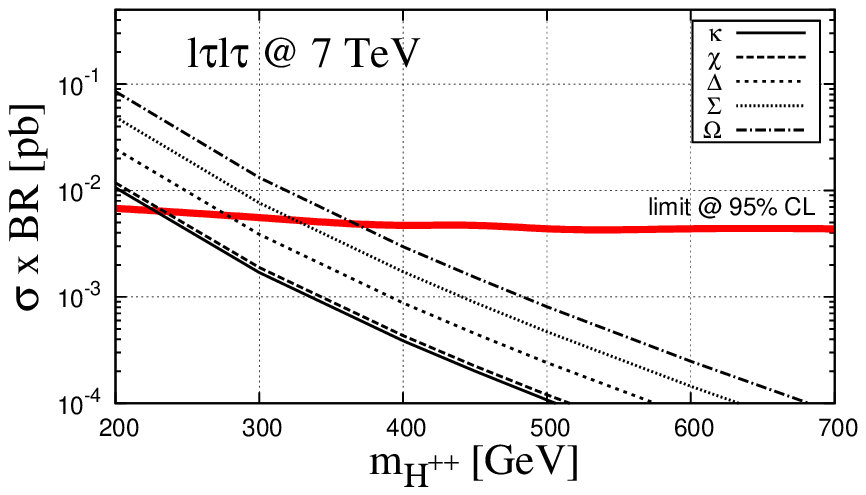} 
& \includegraphics[width=0.49\columnwidth]{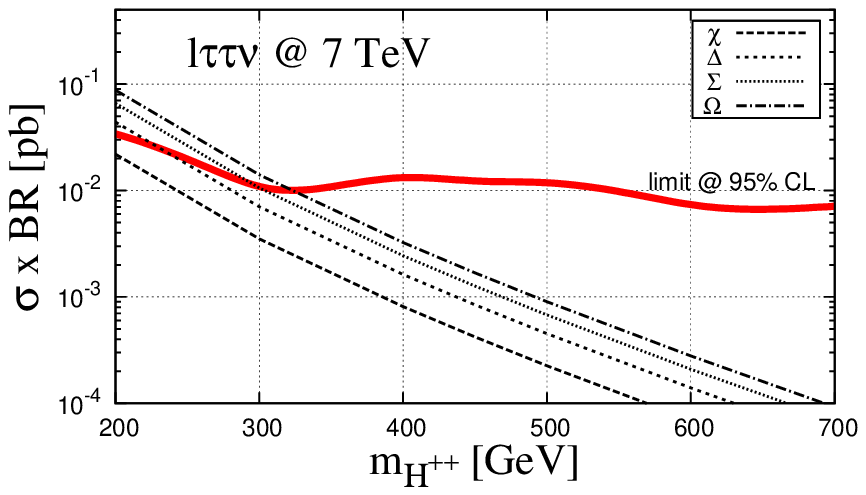} \\ 
\includegraphics[width=0.49\columnwidth]{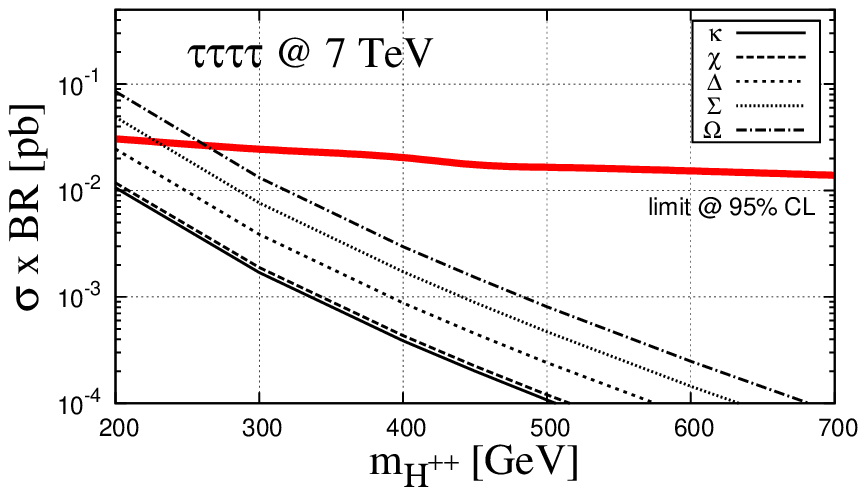} 
& \includegraphics[width=0.49\columnwidth]{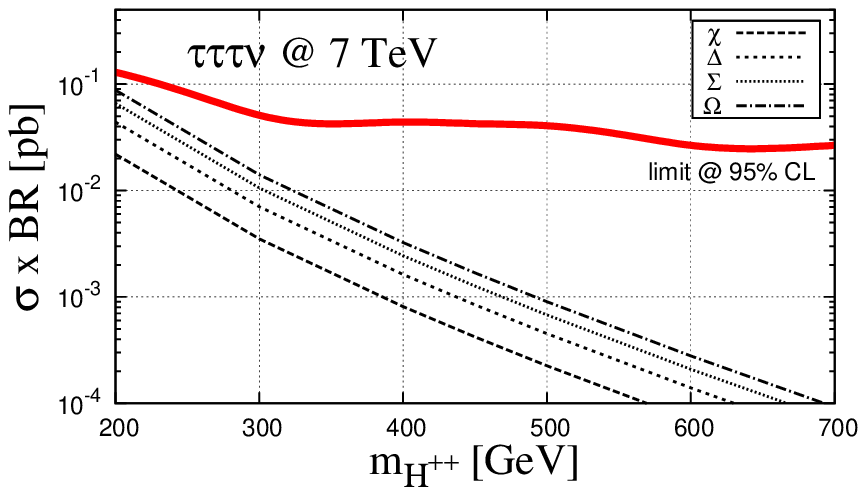} 
\end{tabular}
\caption{Estimated 95 \% C.L. limits on the final modes 
$\ell^\pm \ell^\pm \ell^\mp \ell^\mp$, $\ell^\pm \tau^\pm \ell^\mp \tau^\mp$, 
$\tau^\pm \tau^\pm \tau^\mp \tau^\mp$ (left column from top to bottom) and 
$\ell^\pm \ell^\pm \ell^\mp \nu_\ell$, $\ell^\pm \tau^\pm \ell^\mp \nu_\ell (\tau^\mp \nu_\tau)$, 
$\tau^\pm \tau^\pm \tau^\mp \nu_\tau$ (right column from top to bottom) 
as a function of the doubly-charged scalar mass ${\rm H}^{++}$ 
for $\sqrt s =$ 7 TeV and ${\cal L}_{int} = 4.9$ fb$^{-1}$ at LHC. 
There are superimposed the corresponding cross-sections for the five scalar 
multiplets of lowest isospin and hypercharge containing a doubly-charged component, 
a singlet $\kappa$, a doublet $\chi$, a triplet $\Delta$, a quadruplet $\Sigma$ and a quintuplet $\Omega$.}
\label{fig:7limits}
\end{figure}
The bounds very much coincide with those reported by CMS for a 
doubly-charged scalar mediating the {\it see-saw} of type II, 
ranging from 400 to 200 GeV depending on the scalar decay mode. 
(What in particular implies that the efficiencies we use are consistent  
within the fast simulation algorithm uncertainties with those obtained 
by CMS.) 
However, if the doubly-charged scalar belongs to other type of 
multiplet, its cross-sections vary and so the bounds on its mass. 
We superimpose in the plots the corresponding predictions for 
the five multiplets discussed before, increasing the limits with 
the cross-section (total isospin). The most stringent bounds 
are then obtained for the quintuplet, being typically $\sim 150$ GeV higher  
than for the triplet ({\it see-saw} of type II). 

\begin{figure}[]
\begin{tabular}{cc}
\includegraphics[width=0.49\columnwidth]{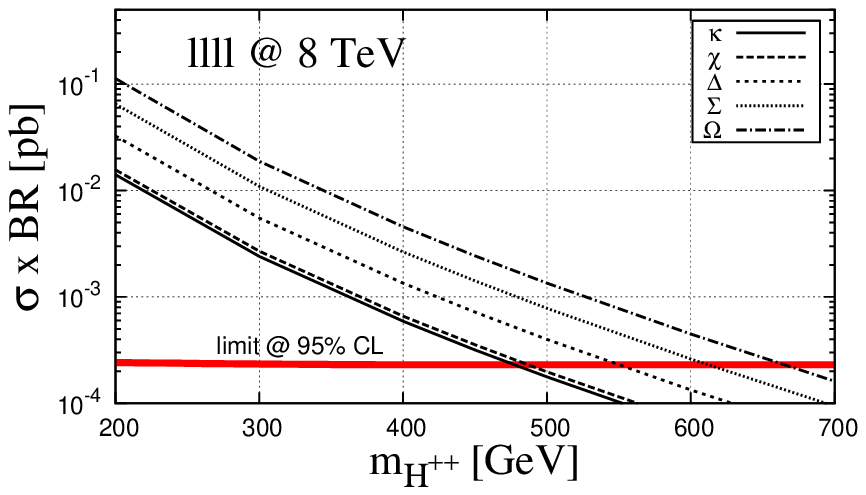} 
& \includegraphics[width=0.49\columnwidth]{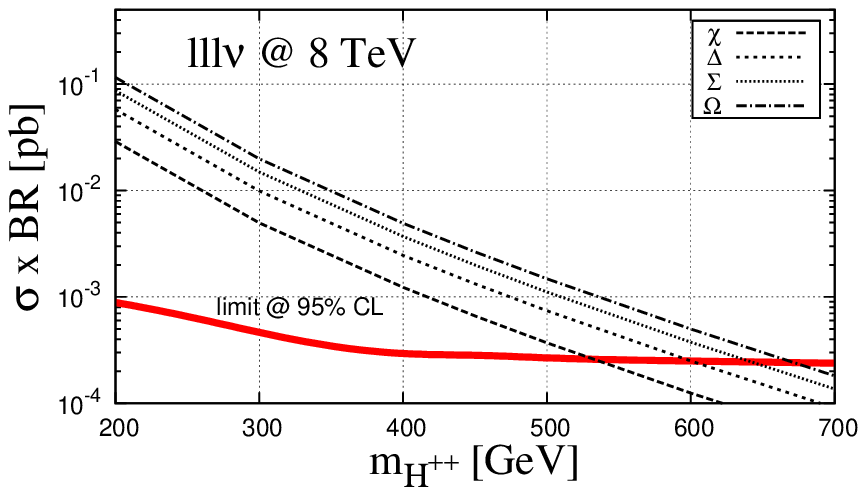} \\ 
\includegraphics[width=0.49\columnwidth]{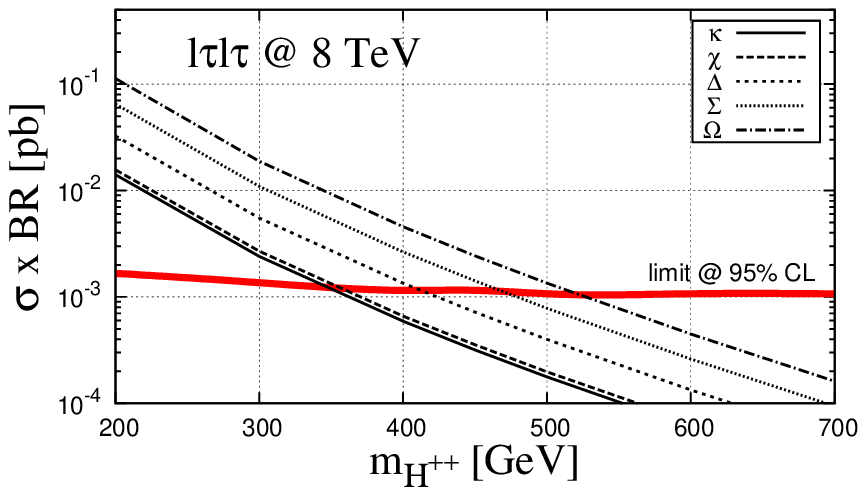} 
& \includegraphics[width=0.49\columnwidth]{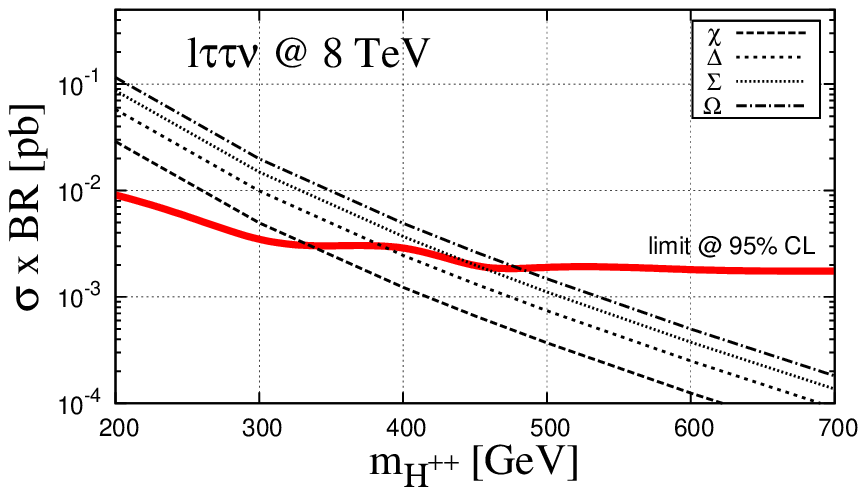} \\ 
\includegraphics[width=0.49\columnwidth]{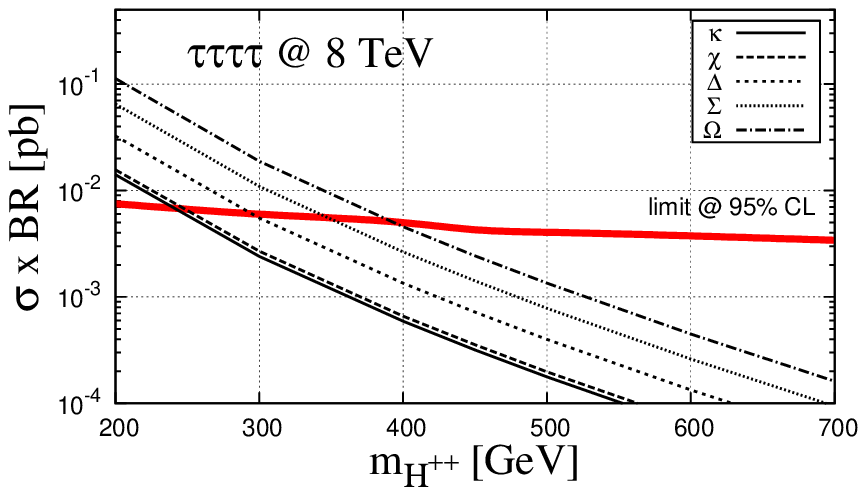} 
& \includegraphics[width=0.49\columnwidth]{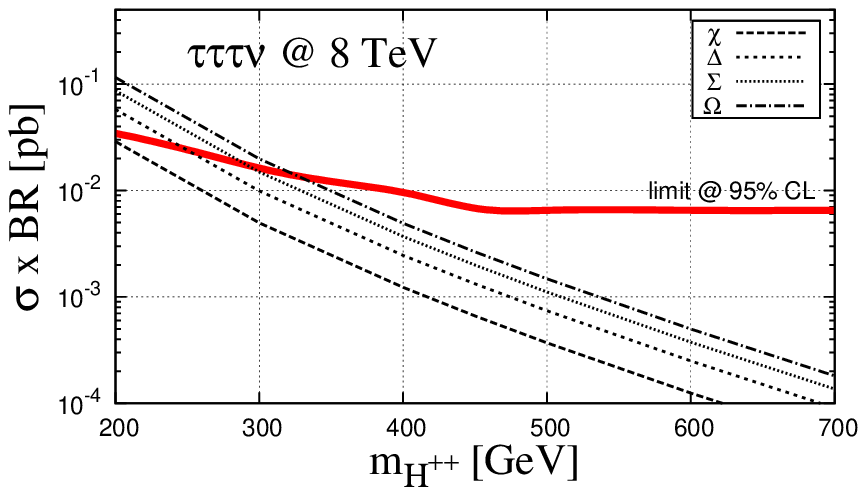} 
\end{tabular}
\caption{The same as in Figure \ref{fig:7limits} but for 8 TeV and 20 fb$^{-1}$.}
\label{fig:8limits}
\end{figure}
Similarly, we can estimate the limits which may be obtained 
at $\sqrt s =$ 8 TeV for an integrated luminosity of 20 fb$^{-1}$, 
assuming that no departure from the SM is observed. 
The efficiencies calculated at 8 TeV for the six decay modes and 
the same CMS cuts are very similar to those at 7 TeV 
for the seven scalar masses.  
In Figure \ref{fig:8limits} we show the corresponding exclusion plots 
\footnote{The number of background and observed events 
at 8 TeV is estimated scaling the 7 TeV values with Eq. (\ref{scaling}). 
Exclusion limits for intermediate masses are obtained by splines interpolation.}. 
Typically, all bounds are around $\sim 100$ GeV higher. 

%
%

\section{Final state-dependent efficiencies and LNV bounds at the LHC}
\label{sec:Efficiencies}

A convenient way of giving a more complete information on the 
experimental bounds on NP is also providing the full set of efficiencies 
for the different processes considered. In this way the limits on 
new models can be in general estimated without performing new analyses. 
For instance, in the case at hand, just giving the bounds on the processes 
with the doubly-charged scalars decaying 100 \% of the time into 
$\ell^\pm \ell^\pm$ 
($pp\rightarrow {\rm H}^{\pm\pm}  {\rm H}^{\mp\mp}\rightarrow \ell^\pm \ell^\pm \ell^\mp \ell^\mp$) 
and into $\ell^\pm \tau^\pm$ 
($pp\rightarrow {\rm H}^{\pm\pm}  {\rm H}^{\mp\mp}\rightarrow \ell^\pm \tau^\pm \ell^\mp \tau^\mp$), 
in turn, 
one can estimate the corresponding limits on a model 
where the doubly-charged scalars decay half of the time into 
each of these two final states, but without being able to 
use the $\ell^\pm \ell^\pm \ell^\mp \tau^\mp$ events 
and hence half of the statistics. 
With this in mind, we collect the efficiencies for the four-lepton and three-lepton analyses in Tables 
\ref{table:EfficienciesFourLeptonsAnalysis} and \ref{table:EfficienciesThreeLeptonsAnalysis}, 
respectively, for all two-body decays of the doubly and singly-charged 
scalars, $\ell\ell, \ell\tau, \tau\tau, WW$ and $\ell\nu, \tau\nu, WZ$, and 
seven scalar masses,  $m_{{\rm H}^{\pm\pm}} = 200$, 300, 400, 450, 500, 600 and 700 GeV.
\begin{table}[] 
\begin{center}
 \begin{adjustwidth}{-0.2cm}{}
{
\begin{tabular}{ l r r r r r r r } 
\ctoprule
$\epsilon^{(4\ell)}_{ij}$ &  $m_{{\rm H}^{\pm\pm}} = 200$ & $\qquad 300$ & $\qquad 400$ 
& $\qquad 450$ & $\qquad 500$ & $\qquad 600$ & $\qquad 700$ \\
\crowcolor 
\cbottomrule
$\ell\ell\ell\ell$ & 53 & 62 & 67 & 68 & 69 & 70 & 71 \\
\ctoprule
$\ell\ell\ell\tau$ & 23 & 27 & 30 & 31 & 32 & 32 & 33 \\
\crowcolor 
\cbottomrule
$\ell\ell\tau\tau$ & 7.2 & 8.7 & 9.4 & 9.4 & 9.5 & 9.7 & 9.9 \\ 
\ctoprule
$\ell\ell WW$ & 0.6 & 1.1 & 1.3 & 1.3 & 1.4 & 1.4 & 1.5 \\
\crowcolor 
\cbottomrule
$\ell\tau \ell\tau$ & 9.0 & 11 & 13 & 14 & 14 & 14 & 14 \\
\ctoprule
$\ell\tau\tau\tau$ & 2.0 & 2.6 & 3.3 & 3.4 & 3.4 & 3.4 & 3.7 \\
\crowcolor 
\cbottomrule
$\ell\tau WW$ & 0.2 & 0.4 & 0.6 & 0.6 & 0.6 & 0.7 & 0.7 \\ 
\ctoprule
$\tau\tau\tau\tau$ & 0.3 & 0.5 & 0.6 & 0.7 & 0.7 & 0.7 & 0.7 \\
\crowcolor 
\cbottomrule 
$\tau\tau WW$ & 0.1 & 0.1 & 0.2 & 0.2 & 0.2 & 0.2 & 0.2 \\ 
\ctoprule
$WWWW$ & $<0.1$ & $<0.1$ & $<0.1$ & $<0.1$ & $<0.1$ & $<0.1$ & $<0.1$ \\
\cbottomrule
\end{tabular}
}
\caption{\label{table:EfficienciesFourLeptonsAnalysis}
Efficiency percentages $\epsilon^{(4\ell)}_{ij}$ for different scalar masses (in GeV) 
and final modes $ij$ for four-lepton analyses at $\sqrt s =$ 7 TeV 
and the $\ell^\pm \tau^\pm \ell^\mp \tau^\mp$ cuts in 
Table \ref{table:EfficienciesOnFourLeptonsAt7TeV}. 
We omit the efficiencies for associated production processes 
because all of them are below $\sim$ 0.1 \%, as these final states do not pass 
the cuts imposed on the four-lepton sample.}
 \end{adjustwidth}
\end{center}
\end{table}
\begin{table}[]
\begin{center}
{
\begin{tabular}{ l r r r r r r r } 
\ctoprule
$\epsilon^{(3\ell)}_{ij}$ &  $m_{{\rm H}^{\pm\pm}} =  200$ & $\qquad 300$ & $\qquad 400$ 
& $\qquad 450$ & $\qquad 500$ & $\qquad 600$ & $\qquad 700$ \\
\crowcolor 
\cbottomrule
$\ell\ell\ell\ell$ & 2.7 & 5.0 & 7.5 & 8.7 & 9.5 & 10 & 11 \\
\ctoprule
$\ell\ell\ell\tau$ & 17 & 25 & 31 & 33 & 34 & 34 & 35 \\
\crowcolor 
\cbottomrule
$\ell\ell\tau\tau$ & 18 & 24 & 28 & 29 & 30 & 31 & 32 \\ 
\ctoprule
$\ell\ell WW$ & 6.9 & 13 & 17 & 18 & 19 & 20 & 21 \\
\crowcolor 
\cbottomrule
$\ell\tau\ell\tau$ & 14 & 19 & 24 & 24 & 25 & 26 & 26 \\
\ctoprule
$\ell\tau\tau\tau$ & 4.9 & 6.9 & 8.6 & 8.6 & 9.0 & 9.2 & 9.3 \\
\crowcolor 
\cbottomrule
$\ell\tau WW$ & 2.3 & 4.6 & 6.3 & 6.6 & 6.7 & 7.0 & 7.2 \\ 
\ctoprule
$\tau\tau\tau\tau$ & 0.8 & 1.2 & 1.4 & 1.4 & 1.4 & 1.6 & 1.6 \\
\crowcolor 
\cbottomrule
$\tau\tau WW$ & 0.4 & 0.8 & 1.0 & 1.0 & 1.1 & 1.2 & 1.2 \\ 
\ctoprule
$WWWW$ & $<0.1$ & 0.3 & 0.4 & 0.4 & 0.5 & 0.5 & 0.5 \\
%
%
\crowcolor 
\cbottomrule
$\ell\ell\ell\nu$ & 38 & 53 & 64 & 66 & 68 & 70 & 72 \\
\ctoprule
$\ell\ell\tau\nu$ & 18 & 26 & 31 & 33 & 34 & 35 & 36 \\
\crowcolor 
\cbottomrule
$\ell\ell WZ$ & 5.0 & 8.5 & 11 & 12 & 13 & 13 & 14 \\
\ctoprule
$\ell\tau \ell\nu$ & 15 & 21 & 26 & 27 & 28 & 29 & 29 \\ 
\crowcolor 
\cbottomrule
$\ell\tau\tau\nu$ & 3.8 & 5.4 & 7.4 & 7.6 & 7.9 & 8.4 & 8.5 \\ 
\ctoprule
$\ell\tau WZ$ & 1.6 & 2.9 & 3.9 & 4.0 & 4.1 & 4.5 & 4.5 \\
\cbottomrule
\end{tabular}
}
\end{center}
\end{table}
\begin{table}[]
\begin{center}
 \begin{adjustwidth}{-0.27cm}{}
{
\begin{tabular}{ l r r r r r r r }
 &  \qquad $\qquad \,\,\,\,\,\,\,\,\,$ & $\qquad \,\,\,\,\,\,\,\,\,$ & $\qquad \,\,\,\,\,\,\,\,\,$  
 & $\qquad \,\,\,\,\,\,\,\,\,$ & $\qquad \,\,\,\,\,\,\,\,\,$ & $\qquad \,\,\,\,\,\,\,\,\,$ & $\qquad \,\,\,\,\,\,\,\,\,$ \\

\crowcolor
\cbottomrule
$\tau\tau \ell\nu$ & 2.3 & 3.4 & 4.3 & 4.5 & 4.6 & 4.8 & 4.9 \\  
\ctoprule
$\tau\tau \tau \nu$ & 0.4 & 0.6 & 0.7 & 0.8 & 0.8 & 0.8 & 0.9 \\ 
\crowcolor 
\cbottomrule
$\tau\tau WZ$ & 0.2 & 0.4 & 0.6 & 0.6 & 0.6 & 0.6 & 0.7 \\  
\ctoprule
$WW\ell\nu$ & 0.3 & 0.7 & 1.0 & 1.0 & 1.0 & 1.1 & 1.2 \\ 
\crowcolor 
\cbottomrule
$WW\tau\nu$ & 0.1 & 0.2 & 0.3 & 0.4 & 0.4 & 0.4 & 0.4 \\  
\ctoprule
$WW WZ$ & $<0.1$ & 0.1 & 0.1 & 0.2 & 0.2 & 0.2 & 0.2 \\ 
\cbottomrule
\end{tabular}
}
\caption{\label{table:EfficienciesThreeLeptonsAnalysis}
Efficiency percentages $\epsilon^{(3\ell)}_{ij}$ for different scalar masses (in GeV) 
and final modes $ij$ for three-lepton analyses at $\sqrt s =$ 7 TeV 
and the $\ell^\pm \tau^\pm \ell^\mp \nu_\tau (\tau^\mp \nu_\ell)$ cuts in 
Table \ref{table:EfficienciesOnThreeLeptonsAt7TeV}.}
 \end{adjustwidth}
\end{center}
\end{table}
As ${\rm H}^{\pm\pm}$ (${\rm H}^{\pm}$) has 4 (3) different two-body decay modes, 
there are a priori $4\times4 + 4\times3 = 28$ final states and hence $\epsilon_{ij}$ 
efficiencies. But for pair production $\epsilon_{ij} = \epsilon_{ji}$, being then only 10 
of the 16 efficiencies independent. In Table \ref{table:EfficienciesFourLeptonsAnalysis} 
we omit the ${\rm H}^{\pm\pm}{\rm H}^{\mp}$ decay modes because all their efficiencies 
are below $\sim$ 0.1 \%.
For both analyses, the applied cuts are common to all final 
states, thus not optimizing the different modes but the full set. 
Following CMS analyses for benchmark points 
we choose the cuts for $\ell^\pm \tau^\pm \ell^\mp \tau^\mp$ in 
Table \ref{table:EfficienciesOnFourLeptonsAt7TeV} and 
for $\ell^\pm \tau^\pm \ell^\mp \nu_\tau (\tau^\mp \nu_\ell)$ in 
Table \ref{table:EfficienciesOnThreeLeptonsAt7TeV} 
to calculate the efficiencies in Tables \ref{table:EfficienciesFourLeptonsAnalysis} 
and \ref{table:EfficienciesThreeLeptonsAnalysis}, respectively.
They grow with the scalar mass because the cuts stay fixed. 
For example, electrons and muons are harder for larger scalar masses 
and hence they satisfy more easily not only the basic cuts but 
the cuts on $\sum p^{\ell}_T$ and on $E^\text{miss}_T$. 
The latter is particularly stringent for pair produced events 
because in this case the missing energy comes either from missed 
leptons or from missing energy measurement errors and 
hence, it is relatively small. In general the missing 
energy and the total transverse momentum in the event 
are correlated, too. 
Whereas the $Z$ veto is also less restrictive for larger masses, 
in contrast with the mass window constraint for events involving tau leptons. 
Changes on parton shower and detector simulation inputs 
stand for variations in the efficiencies of around $\sim$ 15 \%. 
This is the total uncertainty which we assign to the estimates 
in Tables \ref{table:EfficienciesFourLeptonsAnalysis} and 
\ref{table:EfficienciesThreeLeptonsAnalysis}. 
They agree with the efficiencies quoted in \cite{Chatrchyan:2012ya} 
when comparison is possible. 

No dedicated searches for LNV signals have been performed 
in doubly-charged scalar production analyses up to now. 
However, the ATLAS and CMS searches for doubly-charged scalars 
using four and three-lepton samples are also sensitive to LNV final states. 
As already emphasized, we can make use of the 
pertinent efficiencies in Tables \ref{table:EfficienciesFourLeptonsAnalysis} 
and \ref{table:EfficienciesThreeLeptonsAnalysis} in order to derive the 
corresponding bounds, for no event excess has been observed 
in the four and three-lepton analyses. 
We restrict ourselves to the LNV final states $\ell\ell WW$ and $\ell\ell WZ$ because 
they have the largest efficiencies (see Tables \ref{table:EfficienciesFourLeptonsAnalysis} 
and \ref{table:EfficienciesThreeLeptonsAnalysis}), showing the results for the 
three-lepton analysis only for it is by far the most sensitive to LNV. 
For a given integrated luminosity the number of signal events 
$N^{(3\ell)}_{\ell\ell WW (\ell\ell WZ)}$ is estimated 
multiplying ${\mathcal L}_{int}$ by 
the corresponding cross-section, 
$\sigma_{NC (CC)}$ for doubly-charged scalar pair (associated) production, 
times the model branching ratios into $\ell\ell$ and $WW$ ($WZ$) 
times the selection efficiency: 
\begin{align}
N^{(3\ell)}_{\ell\ell WW} & = {\mathcal L}_{int} \times \sigma_{NC} \times \text{BR}({\rm H}^{\pm\pm}\rightarrow \ell^\pm \ell^\pm)
\times\text{BR}({\rm H}^{\mp\mp}\rightarrow W^\mp W^\mp )\times \epsilon^{(3\ell)}_{\ell\ell WW}\ , \nonumber \\
N^{(3\ell)}_{\ell\ell WZ} & = {\mathcal L}_{int} \times \sigma_{CC} \times \text{BR}({\rm H}^{\pm\pm}\rightarrow \ell^\pm \ell^\pm)
\times\text{BR}({\rm H}^{\mp}\rightarrow W^\mp Z ) \times \epsilon^{(3\ell)}_{\ell\ell WZ}\ .
\label{eventnumber}
\end{align}
Making use of these expressions and the number of expected background events and 
observed events we can derive the exclusion plots for $\sigma_{NC (CC)}$ 
(see Appendix \ref{statistics}). 
In Figure \ref{fig:LNVlimits} we plot the corresponding limits assuming that 
the heavy scalars have the same decay rate (50 \%) into light (first two families) lepton 
and gauge boson pairs.  
\begin{figure}[]
\begin{tabular}{cc}
\includegraphics[width=0.49\columnwidth]{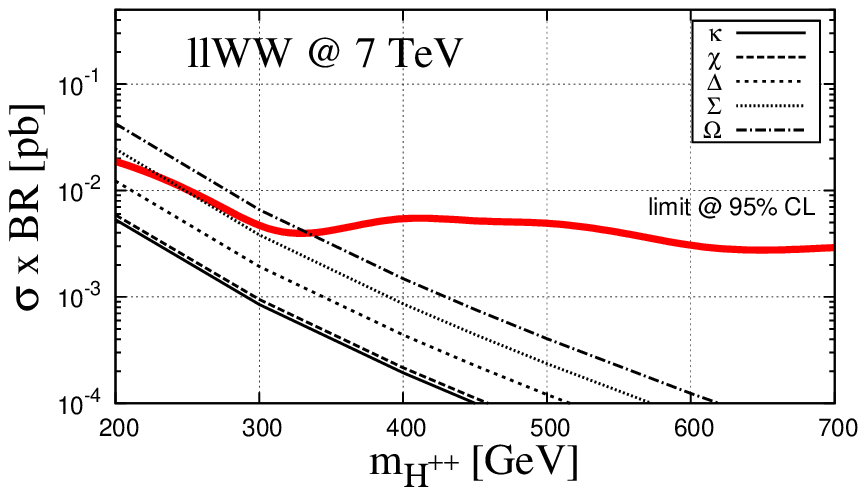} 
& \includegraphics[width=0.49\columnwidth]{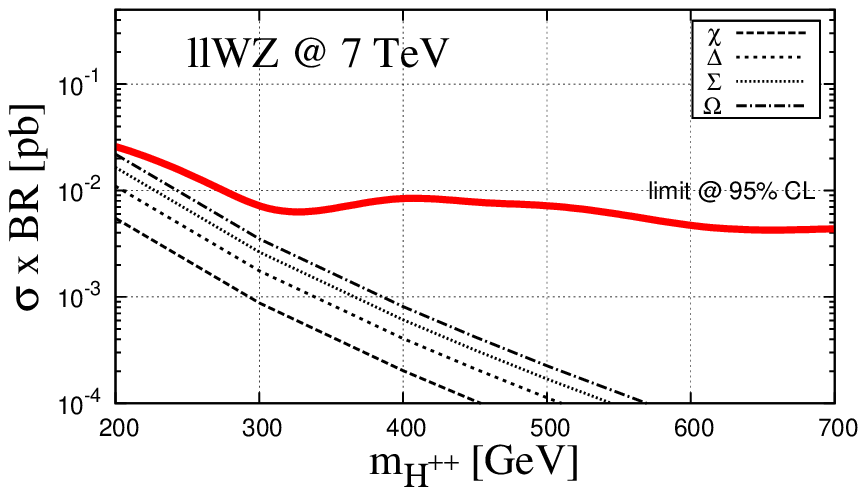} \\ 
\includegraphics[width=0.49\columnwidth]{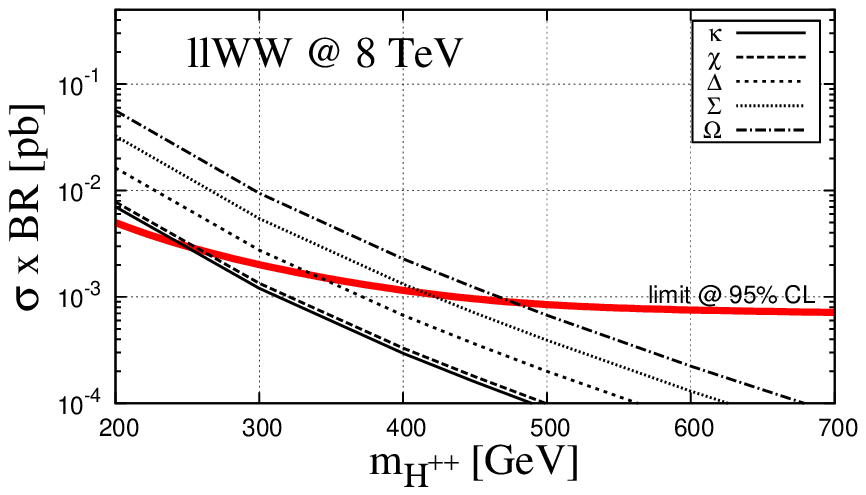} 
& \includegraphics[width=0.49\columnwidth]{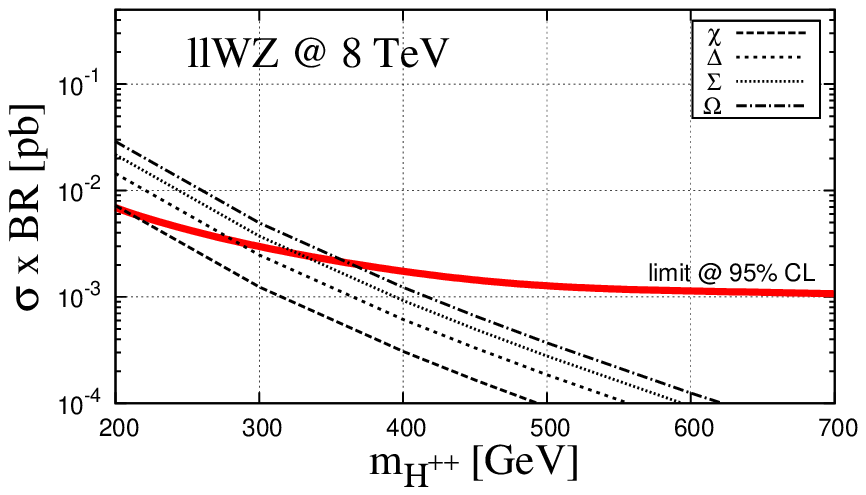} \\ 
\includegraphics[width=0.49\columnwidth]{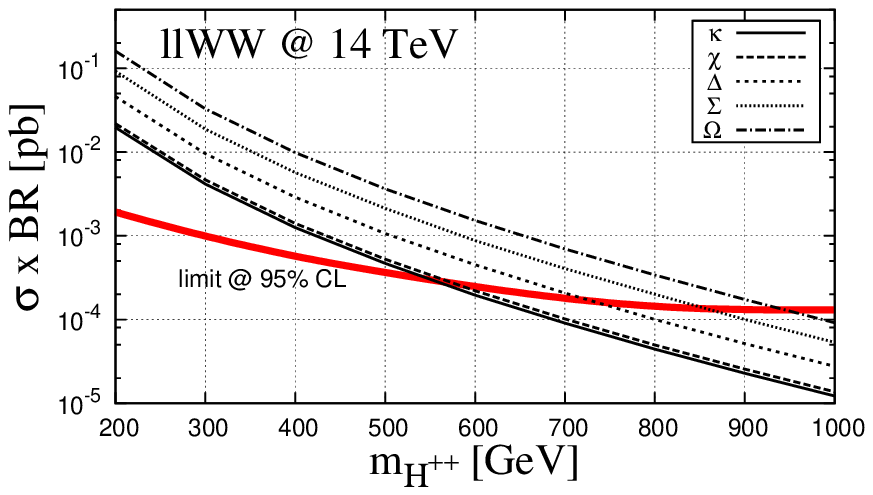} 
& \includegraphics[width=0.49\columnwidth]{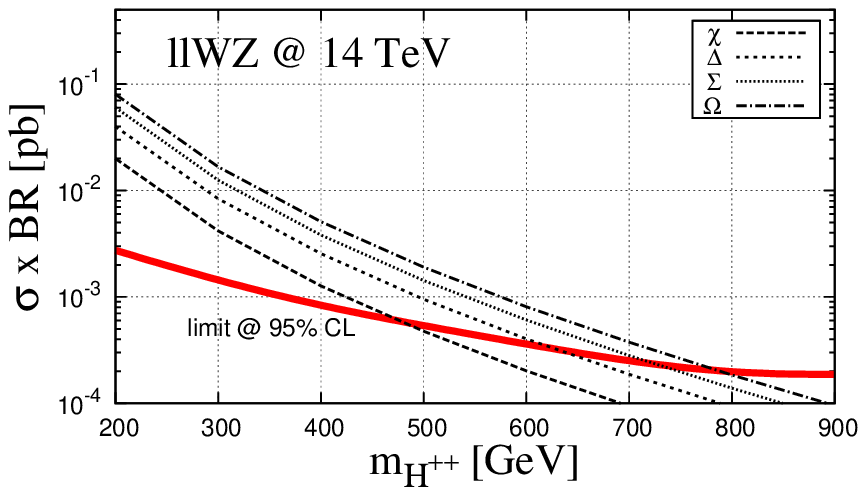} 
\end{tabular}
\caption{95 \% C.L. limits on the LNV channels 
$pp\rightarrow {\rm H}^{\pm\pm}{\rm H}^{\mp\mp} \rightarrow {\ell}^\pm {\ell}^\pm W^\mp W^\mp$ 
(left) and $pp\rightarrow {\rm H}^{\pm\pm}{\rm H}^{\mp} \rightarrow {\ell}^\pm {\ell}^\pm W^\mp Z$ 
(right) as a function of the ${\rm H}^{\pm\pm}$ mass for $\sqrt s =$ 7, 8 and 14 TeV and 
${\cal L}_{int}$ = 4.9, 20 and 100 fb$^{-1}$ at LHC, respectively (from top to bottom). 
There are superimposed the corresponding cross-sections for the five scalar multiplets of 
lowest isospin and hypercharge containing a doubly-charged component: 
the singlet $\kappa$, the doublet $\chi$, the triplet $\Delta$, the quadruplet $\Sigma$ and the 
quintuplet $\Omega$.} 
\label{fig:LNVlimits}
\end{figure}
We superimpose the cross-sections for the different doubly-charged multiplet 
assignments conveniently normalized by the assumed branching ratios: 1/2 for 
pair and 1/4 for associated production.
The exclusion plots for $\ell\ell WW$ ($\ell\ell WZ$) are shown on the 
left (right). 
From top to bottom we gather the LNV bounds at 7, 8 and 14 TeV. 
The number of expected background and observed events for the first two energies 
are the same as for the $\ell^\pm \tau^\pm \ell^\mp \nu_\tau (\tau^\mp \nu_\ell)$ 
three-lepton analysis in the previous section; whereas for 14 TeV we assume 
them to be equal, finding for the backgrounds in Table \ref{table:CrossSections} 
with the same cuts a total of 42, 37, 18, 15, 9, 7 and 0 events 
for an integrated luminosity of 100 fb$^{-1}$ and 
$m_{{\rm H}^{\pm\pm}} =  200$, 300, 400, 450, 500, 600 and 700 GeV, respectively. 
In Tables \ref{table:EfficienciesFourLeptonsAnalyses} and 
\ref{table:EfficienciesThreeLeptonsAnalyses} we collect the efficiencies 
at this energy for four and three-lepton analyses for 
${\rm H}^{\pm\pm} \rightarrow \ell^\pm\ell^\pm, W^\pm W^\pm$ and 
${\rm H}^{\pm} \rightarrow \ell^\pm\nu_\ell, W^\pm Z$ for completeness, 
although only the efficiencies $\epsilon^{(3\ell)}_{\ell\ell WW, \ell\ell WZ}$ for 
the three-lepton analysis (Table \ref{table:EfficienciesThreeLeptonsAnalyses}) 
enter in the calculation of the LNV bounds 
in Figure \ref{fig:LNVlimits}.   
\begin{table}[]
\begin{center}
\begin{adjustwidth}{-0.27cm}{}
{
\begin{tabular}{ l r r r r r r r } 
\ctoprule
$\epsilon^{(4\ell)}_{ij}$ &  $m_{{\rm H}^{\pm\pm}} =  200$ & $\qquad 300$ & $\qquad 400$ 
& $\qquad 450$ & $\qquad 500$ & $\qquad 600$ & $\qquad 700$ \\
\crowcolor 
\cbottomrule
$\ell\ell\ell\ell$ & 54 & 62 & 66 & 68 & 69 & 69 & 70 \\
\ctoprule
$\ell\ell WW$ & 0.6 & 0.9 & 1.1 & 1.4 & 1.4 & 1.4 & 1.5 \\
\crowcolor 
\cbottomrule
$WWWW$ & $<0.1$ & $<0.1$ & $<0.1$ & $<0.1$ & $<0.1$ & $<0.1$ & $<0.1$ \\ 
\ctoprule
$\ell\ell\ell\nu$ & $<0.1$ & $<0.1$ & $<0.1$ & $<0.1$ & $<0.1$ & $<0.1$ & $<0.1$ \\ 
\crowcolor 
\cbottomrule 
$\ell\ell WZ$ & $<0.1$ & 0.1 & 0.1 & 0.1 & 0.1 & 0.1 & 0.1 \\
\ctoprule
$WW\ell\nu$ & $<0.1$ & $<0.1$ & $<0.1$ & $<0.1$ & $<0.1$ & $<0.1$ & $<0.1$ \\
\crowcolor 
\cbottomrule 
$WW WZ$ & $<0.1$ & $<0.1$ & $<0.1$ & $<0.1$ & $<0.1$ & $<0.1$ & $<0.1$ \\
\ctoprule
\end{tabular}
}
\caption{\label{table:EfficienciesFourLeptonsAnalyses}
Efficiency percentages $\epsilon^{(4\ell)}_{ij}$ for different scalar masses (in GeV) 
and final modes $ij$ for four-lepton analyses at $\sqrt s =$ 14 TeV 
and the $\ell^\pm \tau^\pm \ell^\mp \tau^\mp$ cuts in 
Table \ref{table:EfficienciesOnFourLeptonsAt7TeV}.}
 \end{adjustwidth}
\end{center}
\end{table}
\begin{table}[]
\begin{center}
 \begin{adjustwidth}{-0.25cm}{}
{
\begin{tabular}{ l r r r r r r r } 
\ctoprule
$\epsilon^{(3\ell)}_{ij}$ &  $m_{{\rm H}^{\pm\pm}} =  200$ & $\qquad 300$ & $\qquad 400$ 
& $\qquad 450$ & $\qquad 500$ & $\qquad 600$ & $\qquad 700$ \\
\crowcolor 
\cbottomrule
$\ell\ell\ell\ell$ & 3.9 & 6.9 & 10 & 11 & 12 & 13 & 14 \\
\ctoprule
$\ell\ell WW$ & 7.4 & 14 & 18 & 20 & 21 & 22 & 23 \\
\crowcolor 
\cbottomrule
$WWWW$ & $<0.1$ & 0.3 & 0.4 & 0.5 & 0.6 & 0.7 & 0.7 \\ 
\ctoprule
$\ell\ell\ell\nu$ & 40 & 55 & 66 & 68 & 70 & 72 & 74 \\ 
\crowcolor 
\cbottomrule
$\ell\ell WZ$ & 5.2 & 9.3 & 13 & 13 & 14 & 15 & 16 \\
\ctoprule
$WW\ell\nu$ & 0.3 & 0.8 & 1.0 & 1.1 & 1.1 & 1.2 & 1.3 \\
\crowcolor 
\cbottomrule
$WW WZ$ & $<0.1$ & 0.2 & 0.2 & 0.3 & 0.2 & 0.3 &  0.3\\
\ctoprule
\end{tabular}
}
\caption{\label{table:EfficienciesThreeLeptonsAnalyses}
Efficiency percentages $\epsilon^{(3\ell)}_{ij}$ for different scalar masses (in GeV) 
and final modes $ij$ for three-lepton analyses at $\sqrt s =$ 14 TeV 
and the $\ell^\pm \tau^\pm \ell^\mp \nu_\tau (\tau^\mp \nu_\ell)$ cuts in 
Table \ref{table:EfficienciesOnThreeLeptonsAt7TeV}.}
 \end{adjustwidth}
\end{center}
\end{table}

At 7 TeV and with an integrated luminosity ${\cal L}_{int} =$ 4.9 fb$^{-1}$ 
LHC has in general no sensitivity to the LNV signals considered. 
But the expected bounds at 8 TeV with ${\cal L}_{int} =$ 20 fb$^{-1}$ 
range from $\sim 200$ to $500$ GeV depending on the scalar multiplet. 
These limits can be up to $\sim 500$ GeV larger at 14 TeV with an 
integrated luminosity of 100 fb$^{-1}$. 
Several comments are in order, however: 
({\it i}) These estimates will be improved by the experimental 
collaborations when cuts and efficiencies are optimized 
for these searches. 
({\it ii}) LNV processes are in general rare 
and as previously emphasized, only in special regions in parameter 
space they are relatively large with almost half of the events 
from doubly-charged scalar pair and associated production 
violating LN. This is what we have assumed to draw Figure \ref{fig:LNVlimits}. 
If the branching ratios are different we have to normalize the 
H cross-sections accordingly to read the corresponding limit from the Figure. 
({\it iii}) Relatively large LNV signals are more natural in more elaborated models, as 
for example, those with neutrino masses generated radiatively 
\cite{delAguila:2011gr,delAguila:2012nu,Gustafsson:2012vj,Babu:2009aq}.

Two last comments are in order: \hfill\break 
({\it i}) LNV analyses must be improved by experimentalists 
not only doing a better job using real data but adapting the searches (cuts) 
to the signal characteristics. \hfill\break  
({\it ii}) On the other hand, if no departure from the SM predictions 
is observed, one can use all final modes to constrain the model, 
independently of whether LN is or is not violated or the production mechanism. 
The number of signal events is in this case 
\begin{align}
N^{(A)} = {\mathcal L}_{int} \times & \left[ \sigma_{NC} \times 
\sum_{ij} \text{BR}({\rm H}^{\pm\pm}\rightarrow i)
\times\text{BR}({\rm H}^{\mp\mp}\rightarrow j)\times \epsilon^{(A)}_{ij} \right. \nonumber \\
& + \left. \sigma_{CC} \times \sum_{rs} \text{BR}({\rm H}^{\pm\pm}\rightarrow r)
\times\text{BR}({\rm H}^{\mp}\rightarrow s) \times \epsilon^{(A)}_{rs} \right] \ ,
\label{eventnumber}
\end{align}
where the sum is over all ${\rm H}^{\pm\pm}$ ($i,j,r$) and ${\rm H}^{\mp}$ ($s$) 
final states contributing to analysis $A$. 
Here $A = 4\ell , 3\ell$. In general, all analyses can be also taken into 
account together when deriving generic limits on a given model 
using the CL$_s$ method by assigning each analysis to a 
different bin. In our case this means to 2 bins in order to account for the $4\ell$ and $3\ell$ analyses.  
 
%
%

%
\section{Conclusions}
\label{sec:concl}

If neutrino masses are Majorana, LN must be violated at some scale 
\footnote{Independently of whether it is gauged \cite{Duerr:2013dza} or not.}. 
The important question at the LHC era is if it is broken at the TeV scale and 
hence, if LNV can be observed at the LHC. 
Among the simplest SM extensions which can give neutrinos a mass 
and predict new resonances at the LHC reach, 
the {\it see-saw} of type II provides the cleanest signal: 
in this case the mediator is a heavy scalar triplet of hypercharge 1, 
($\Delta^{++}, \Delta^{+}, \Delta^{0}$), which can resonate in the same-sign 
dilepton channel, $\Delta^{++} \rightarrow l^\pm l^\pm$. 
The production rate is of EW size, which is the largest possible because 
a priori this NP has no color, and the decay products 
can be isolated electrons and muons with large momenta and missing energy. 
As these scalars have non-zero LN (equal to 2) and LN can be only very tiny broken 
in the SM sector, they must be pair produced. Thus, whatever the decays 
are, the final state must have at least four fermions typically carrying 
each of them one quarter of the total available energy.  
   
In this paper we have extended the {\it see-saw} of type II and classified the 
scalar multiplets {\rm H} which produce the same signals, 
paying special attention to their LNV decays, 
$pp\rightarrow {\rm H}^{\pm\pm}{\rm H}^{\mp\mp} \rightarrow {\ell}^\pm {\ell}^\pm W^\mp W^\mp$ and 
$pp\rightarrow {\rm H}^{\pm\pm}{\rm H}^{\mp} \rightarrow {\ell}^\pm {\ell}^\pm W^\mp Z$, 
not explicitly considered up to now. 
All those multiplets include doubly-charged scalars, 
being then possible to characterize {\rm H} by the production and decay of their doubly-charged component. 
In particular, we have discussed the main doubly-charged scalar production mechanisms and 
worked out the corresponding Feynman rules in detail, providing a 
\texttt{MadGraph5} model for Monte Carlo simulations. 
Finally, as a practical application we have reproduced the current searches 
for doubly-charged scalars by CMS and ATLAS at 
$\sqrt s = 7$ TeV with ${\mathcal L}_{int} = 4.9$ fb$^{-1}$ \cite{Chatrchyan:2012ya,ATLAS:2012hi}. 
Present limits on their mass can be as large as 400 GeV 
for the scalar triplet mediating the {\it see-saw} of type II, depending on the decay mode. 
These bounds raise up to 500 GeV for the scalar multiplet of highest 
isospin which we have worked out, a quintuplet of hypercharge 0, 
($\Omega^{++}, \Omega^{+}, \Omega^{0}, \Omega^{-}, \Omega^{--}$).
Using similar cuts we have also estimated the expected bounds at 8 TeV if no 
event excess is observed with an integrated 
luminosity of 20 fb$^{-1}$, being typically 150 GeV larger than those obtained at 7 TeV. 
These bounds can be translated to any general model if 
the efficiencies for the relevant (all) channels are known. 
We have provided 
a table of Monte Carlo estimates for these efficiencies, which 
we have then used to estimate 
the bounds on LNV for different LHC runs. 
These limits 
can be near the TeV for the most favorable 
case at 14 TeV with an integrated luminosity of 
100 fb$^{-1}$ (see Figure \ref{fig:LNVlimits}). 
However, they are only significant if the doubly-charged scalars have similar 
decay rates into same-sign lepton and boson pairs, 
as it is more natural in less simple models 
\cite{delAguila:2011gr,delAguila:2012nu,Gustafsson:2012vj,Babu:2009aq}. At any rate, it must be emphasized that the analyses we have performed to obtain these bounds are not optimized for LNV searches. As a matter of fact, we have used the only analyses sensitive to doubly-charged scalar production experimentally available up to now. Hence, more sophisticated analyses taking into account the specific topology of LNV processes would have to be performed in order to extract all the information from future runs. A first attempt in this direction was given in \cite{delAguila:2013yaa,delAguila:2013hla}.

We have assumed the most optimistic scenario 
in order to estimate the LHC potential for LNV searches. 
We not only assume that doubly-charged scalars have similar 
decay rates into same-sign lepton and gauge boson pairs, but that 
cascade decays within the multiplet are negligible 
(see footnotes \ref{mixing} and \ref{triply} for further comments). Which requires 
that the scalar mixing is rather small 
\cite{Grifols:1989xe,Perez:2008ha}. 
It is thus worth to work out specific models where the optimal scenario 
adopted in our phenomenological approach is naturally realized. 

\section*{Acknowledgements}

We thank A. Santamaria and J. Wudka for previous collaboration and a careful reading of the manuscript, 
and J. Alcaraz and J. Santiago for useful comments. 
This work has been supported in part by the Ministry of Economy and 
Competitiveness (MINECO), grant FPA2010-17915, and by the Junta de Andaluc{\'\i}a, 
grants FQM 101 and FQM 6552. M.C. is supported by the MINECO under the FPU program.

\appendix
\section{Contact interactions}
\label{effectiveproduction}

There can be also contact interactions generated by (even) heavier 
particles after integrating them out. The corresponding operators, 
however, are at least of dimension 6 and hence suppressed by two 
powers of the heavier effective scale $\Lambda$. What stands for  
a suppression of the doubly-charged scalar production cross-section 
at LHC of the order of $(m_{{\rm H}^{++}} / \Lambda)^4$. 

Indeed, the contact interactions of lowest dimension must involve 
the gauge invariant contraction of two gluon field strength tensors 
$G_a^{\mu\nu} G^a_{\mu\nu}$ or a colorless quark bilinear 
$\overline{Q} Q'$. 
In the former case the EW singlet of lowest dimension involving at 
least one scalar multiplet H with a doubly-charged component ${\rm H}^{++}$
is ${\rm H}^\dagger {\rm H}$. Thus, the corresponding lowest order operator 
$G_a^{\mu\nu} G^a_{\mu\nu} {\rm H}^\dagger {\rm H}$ 
is of dimension 6, being also the only one of this dimension and form. 

On the other hand, there are two possible types of quark bilinears 
depending on the fermion chirality:  $\overline{Q_{L(R)}} \gamma_\mu Q'_{L(R)}$ 
and $\overline{Q_{L(R)}} Q'_{R(L)}$, where $Q^{(\prime)}_L = q_L$ is 
the left-handed quark doublet and $Q^{(\prime)}_R = u_R, d_R$ are 
the corresponding right-handed singlets. 
But there is no invariant product of any of them with only 
one scalar multiplet H coupling the quark bilinear to 
the H doubly-charged component in the unitary gauge, 
and hence contributing to doubly-charged scalar production 
at hadron colliders. 
Besides, the vector quark bilinear requires an additional covariant derivative 
to ensure that the operator is Lorentz invariant. 
Then, in this case there is only one invariant operator of lowest 
dimension involving two H multiplets and not suppressed by a small quark 
Yukawa coupling: 
${\rm H}^\dagger (D^\mu {\rm H}) \overline{Q_{L(R)}}\gamma_\mu Q'_{L(R)}$. 
Other operators of dimension 6 with the covariant derivative acting on 
the quark fields can be shown to be suppressed by a small quark 
Yukawa coupling using the equations of motion; while the operator 
with the covariant derivative acting on the other scalar multiplet 
can be written as a combination of all the other operators integrating 
by parts. 
 
The lowest order operators involving the other quark bilinear and 
two H multiplets are also of dimension 6 because they must involve 
at least a $\phi$ factor to render the operator invariant 
under isospin transformations. 
Thus, all such operators resemble 
$\overline{q}_{L} \phi u_{R} {\rm H}^{\dagger} {\rm H}$, being hence suppressed 
after EWSB by a $v/ m_{{\rm H}^{++}}$ factor relative to 
${\rm H}^\dagger (D^\mu {\rm H}) \overline{Q_{L(R)}}\gamma_\mu Q'_{L(R)}$. 
In summary, the largest contribution (at least formally) results from 
the operators of this form which can be, 
for instance, obtained after integrating out a heavy $Z'$. 
But their contribution is in general suppressed far away from the heavy 
resonance by a $( m_{{\rm H}^{++}} / \Lambda )^2$ factor 
(as it is the case for $G_a^{\mu\nu} G^a_{\mu\nu} {\rm H}^\dagger {\rm H}$, too). 

\section{Simulation and analyses}
\label{sec:SimulationAndAnalyses}

As indicated in the text, the signal is simulated using 
\texttt{MadGraph5} \cite{Alwall:2011uj} supplemented with the corresponding 
\texttt{UFO} model including the scalar interactions 
(the scalar gauge couplings $\mathcal{SSV}$ and $\mathcal{SSVV}$ in Eq. (\ref{lagrangian}), 
and the scalar couplings to fermions $\mathcal{SFF}$ and to gauge bosons $\mathcal{SVV}$ 
in Eqs. (\ref{eq:yuk}) and (\ref{WWH}), respectively). 
The \texttt{UFO} model can be found in
\href{http://cafpe.ugr.es/index.php/pages/other/software}
{\textsf{http://cafpe.ugr.es/index.php/pages/other/software}} 
in the package \textsf{LNV-Scalars\_UFO.tar.gz}. 
The doubly (singly) charged scalar components 
for the singlet, doublet, triplet, quadruplet and quintuplet 
are identified by \texttt{hs2, hd2, ht2, hq2} and \texttt{hk2} 
(\footnote{Since the singlet has no singly-charged component, 
no name is assigned for this case.}
\texttt{hd1, ht1, hq1} and \texttt{hk1}), respectively. 
In addition, the package contains a set of \texttt{Param Cards} 
for the seven masses considered in our analysis, 
as well as a README file with examples for the
production of doubly-charged scalars belonging to different multiplets 
and for several processes.

\section{CL$_s$ method}
\label{statistics}

We use the CL$_s$ method for the calculation of the exclusion limits \cite{Read:2002hq}. 
This method associates to a sample with N bins the statistic  
\begin{align}
Q &= \prod_i \frac{(s_i+b_i)^{\tilde{n}_i}e^{-(s_i+b_i)}}{b_i^{\tilde{n}_i}e^{-b_i}} \nonumber \\
  &= e^{-\sum_i s_i}\prod_i\left[1+\frac{s_i}{b_i}\right]^{\tilde{n}_i}\ ,
\end{align}
where $b_i$ and $s_i$ are the number of predicted background events and 
of expected signal events for bin $i$, respectively, and $\tilde{n}_i$ is the 
Poisson-distributed variable with mean $s_i+b_i$ ($b_i$) for the 
signal+background (background-only) hypothesis. 
The confidence estimators 
\begin{align}
\text{CL}_{s+b} &= 1-\int_{Q_{obs}}^\infty P_{s+b}(Q) dQ \quad {\rm and} \nonumber \\ 
\text{CL}_b &= 1 - \int_{Q_{obs}}^\infty P_b(Q) dQ 
\end{align}
are then defined integrating the corresponding density functions $P_{s+b}(Q)$ and $P_b(Q)$, 
respectively, up to $Q_{obs}$, which is the $Q$ value for $\tilde{n}_i$ equal to the number 
of observed events $n_i$. 
Thus, parameter space regions excluded at the 95 \% confidence level (C.L.) 
can be obtained requiring that 
CL$_s \equiv \text{CL}_{s+b}/\text{CL}_b \le 0.05$. 

Either $Q$ or $\log{Q}$ can be used as statistic, although the latter is more convenient for 
calculating CL$_s$ if there is only one bin (counting experiment). 
In this case, 
\begin{equation}
Q = e^{-s}\left(1+\frac{s}{b}\right)^{\tilde{n}} \Rightarrow \log{Q} = -s + \tilde{n}\left(1+\frac{s}{b}\right) .
\end{equation}
Hence, $\log{Q}$ is distributed as $\tilde{n}$ up to a scale factor and a shift. 
But none of them changes the ratio of areas defining CL$_s$, being then 
easier to use the $\tilde{n}$ distribution as statistic.

\bibliographystyle{JHEP}

\begin{thebibliography}{99}
%


\bibitem{Aad:2012tfa}
  G.~Aad {\it et al.}  [ATLAS Collaboration],
  Phys.\ Lett.\ B {\bf 716} (2012) 1
  [arXiv:1207.7214 [hep-ex]]; 
  Phys.\ Lett.\ B {\bf 726} (2013) 120
  [arXiv:1307.1432 [hep-ex]].

\bibitem{Chatrchyan:2012ufa}
  S.~Chatrchyan {\it et al.}  [CMS Collaboration],
  Phys.\ Lett.\ B {\bf 716} (2012) 30
  [arXiv:1207.7235 [hep-ex]]; 
  Phys.\ Rev.\ Lett.\  {\bf 110} (2013) 081803
  [arXiv:1212.6639 [hep-ex]].


\bibitem{Giardino:2013bma}
  P.~P.~Giardino, K.~Kannike, I.~Masina, M.~Raidal and A.~Strumia,
  arXiv:1303.3570 [hep-ph]; 
  J.~Ellis and T.~You,
  JHEP {\bf 1306} (2013) 103
  [arXiv:1303.3879 [hep-ph]]; 
  A.~Djouadi and G.~Žg.~Moreau,
  arXiv:1303.6591 [hep-ph]; 
  G.~Belanger, B.~Dumont, U.~Ellwanger, J.~F.~Gunion and S.~Kraml,
  Phys.\ Rev.\ D {\bf 88} (2013) 075008
  [arXiv:1306.2941 [hep-ph]]; 
  S.~Heinemeyer {\it et al.}  [LHC Higgs Cross Section Working Group Collaboration],
  arXiv:1307.1347 [hep-ph]; and references there in.


\bibitem{Barbieri:2000gf}
  R.~Barbieri and A.~Strumia,
  hep-ph/0007265.
  
\bibitem{delAguila:2011zs}
  F.~del Aguila and J.~de Blas,
  Fortsch.\ Phys.\  {\bf 59} (2011) 1036
  [arXiv:1105.6103 [hep-ph]].


\bibitem{Goebel:2010ux}
  M.~Goebel [Gfitter Group Collaboration],
  PoS ICHEP {\bf 2010} (2010) 570
  [arXiv:1012.1331 [hep-ph]]; 
  [ALEPH and CDF and D0 and DELPHI and L3 and OPAL and SLD and LEP Electroweak Working Group and Tevatron Electroweak Working Group and SLD Electroweak and Heavy Flavour Groups Collaborations],
  arXiv:1012.2367 [hep-ex]. 


\bibitem{Baak:2012kk}
  M.~Baak, M.~Goebel, J.~Haller, A.~Hoecker, D.~Kennedy, R.~Kogler, K.~Moenig and M.~Schott {\it et al.},
  Eur.\ Phys.\ J.\ C {\bf 72} (2012) 2205
  [arXiv:1209.2716 [hep-ph]]; 
  A.~Falkowski, F.~Riva and A.~Urbano,
  arXiv:1303.1812 [hep-ph].
  
\bibitem{deBlas:2013gla}
  J.~de Blas,
  arXiv:1307.6173 [hep-ph].

 
\bibitem{Beringer:1900zz}
  J.~Beringer {\it et al.}  [Particle Data Group Collaboration],
  Phys.\ Rev.\ D {\bf 86} (2012) 010001.
  

\bibitem{Mohapatra:1998rq}
  R.~N.~Mohapatra and P.~B.~Pal,
  World Sci.\ Lect.\ Notes Phys.\  {\bf 60} (1998) 1
   [World Sci.\ Lect.\ Notes Phys.\  {\bf 72} (2004) 1].

  
\bibitem{Weinberg:1979sa}
S.~Weinberg, 
{Phys.  Rev. Lett.} {\bf 43} (1979) 1566--1570.


\bibitem{Furry:1939qr}
  W.~H.~Furry,
  Phys.\ Rev.\  {\bf 56} (1939) 1184.
  
\bibitem{Vergados:2002pv}
  J.~D.~Vergados,
  Phys.\ Rept.\  {\bf 361} (2002) 1
  [hep-ph/0209347].

\bibitem{Avignone:2007fu}
  F.~T.~Avignone, III, S.~R.~Elliott and J.~Engel,
  Rev.\ Mod.\ Phys.\  {\bf 80} (2008) 481
  [arXiv:0708.1033 [nucl-ex]]; 
A.~S. Barabash, {\it {Double beta decay experiments}},  {\it Phys. Part. Nucl.}
  {\bf 42} (2011) 613--627, [arXiv:1107.5663 [nucl-ex]].


\bibitem{Schechter:1981bd}
  J.~Schechter and J.~W.~F.~Valle,
  Phys.\ Rev.\ D {\bf 25} (1982) 2951; 
  K.~S.~Babu and C.~N.~Leung,
  Nucl.\ Phys.\ B {\bf 619} (2001) 667
  [hep-ph/0106054]; 
  K.~-w.~Choi, K.~S.~Jeong and W.~Y.~Song,
  Phys.\ Rev.\ D {\bf 66} (2002) 093007
  [hep-ph/0207180]; 
  J.~Engel and P.~Vogel,
  Phys.\ Rev.\ C {\bf 69} (2004) 034304
  [nucl-th/0311072].
    
\bibitem{de Gouvea:2007xp}
  A.~de Gouvea and J.~Jenkins,
  Phys.\ Rev.\ D {\bf 77} (2008) 013008
  [arXiv:0708.1344 [hep-ph]].
  
\bibitem{delAguila:2011gr}
  F.~del Aguila, A.~Aparici, S.~Bhattacharya, A.~Santamaria and J.~Wudka,
  JHEP {\bf 1205} (2012) 133
  [arXiv:1111.6960 [hep-ph]].  

\bibitem{delAguila:2012nu}
  F.~del Aguila, A.~Aparici, S.~Bhattacharya, A.~Santamaria and J.~Wudka,
  JHEP {\bf 1206} (2012) 146
  [arXiv:1204.5986 [hep-ph]]; 
  PoS Corfu {\bf 2012} (2013) 028
  [arXiv:1305.4900 [hep-ph]].

\bibitem{Gustafsson:2012vj}
  M.~Gustafsson, J.~M.~No and M.~A.~Rivera,
  Phys.\ Rev.\ Lett.\  {\bf 110} (2013) 211802
  [arXiv:1212.4806 [hep-ph]].
  
\bibitem{Franceschini:2013aha}
  R.~Franceschini and R.~N.~Mohapatra,
  arXiv:1306.6108 [hep-ph].
  

\bibitem{Keung:1983uu}
  W.~-Y.~Keung and G.~Senjanovic,
  Phys.\ Rev.\ Lett.\  {\bf 50} (1983) 1427.


\bibitem{Fukugita:1986hr}
  M.~Fukugita and T.~Yanagida,
  Phys.\ Lett.\ B {\bf 174} (1986) 45.


\bibitem{Davidson:2008bu}
  S.~Davidson, E.~Nardi and Y.~Nir,
  Phys.\ Rept.\  {\bf 466} (2008) 105
  [arXiv:0802.2962 [hep-ph]]; 
  G.~C.~Branco, R.~G.~Felipe and F.~R.~Joaquim,
  Rev.\ Mod.\ Phys.\  {\bf 84} (2012) 515
  [arXiv:1111.5332 [hep-ph]]; 
  T.~Hambye,
  New J.\ Phys.\  {\bf 14} (2012) 125014
  [arXiv:1212.2888 [hep-ph]].

  
\bibitem{Minkowski:1977sc}
  P.~Minkowski,
  Phys.\ Lett.\ B {\bf 67} (1977) 421; 
  T.~Yanagida,
  Conf.\ Proc.\ C {\bf 7902131} (1979) 95; 
  M.~Gell-Mann, P.~Ramond and R.~Slansky,
  Conf.\ Proc.\ C {\bf 790927} (1979) 315
  [arXiv:1306.4669 [hep-th]]; 
  S.~L.~Glashow,
  NATO Adv.\ Study Inst.\ Ser.\ B Phys.\  {\bf 59} (1980) 687; 
  R.~N.~Mohapatra and G.~Senjanovic,
  Phys.\ Rev.\ Lett.\  {\bf 44} (1980) 912.


\bibitem{Schechter:1980gr}
  J.~Schechter and J.~W.~F.~Valle,
  Phys.\ Rev.\ D {\bf 22} (1980) 2227; 
  M.~Magg and C.~Wetterich,
  Phys.\ Lett. {\bf B94} (1980) 61; 
  T.~P.~Cheng and L.~F.~Li,
  Phys.\ Rev. {\bf D22} (1980) 2860; 
  G.~B.~Gelmini and M.~Roncadelli,
  Phys.\ Lett. {\bf B99} (1981) 411; 
  G.~Lazarides, Q.~Shafi and C.~Wetterich,
  Nucl.\ Phys. {\bf B181} (1981) 287; 
  R.~N.~Mohapatra and G.~Senjanovic,
  Phys.\ Rev. {\bf D23} (1981) 165.


\bibitem{Foot:1988aq}
  R.~Foot, H.~Lew, X.~G.~He and G.~C.~Joshi,
  Z.\ Phys. {\bf C44} (1989) 441; 
  E.~Ma,
  Phys.\ Rev.\ Lett. {\bf 81} (1998) 1171 
  [hep-ph/9805219].

  
\bibitem{Hektor:2007uu}
  A.~Hektor, M.~Kadastik, M.~Muntel, M.~Raidal and L.~Rebane,
  Nucl.\ Phys.\ B {\bf 787} (2007) 198
  [arXiv:0705.1495 [hep-ph]].
 
\bibitem{delAguila:2008cj}
F.~del Aguila and J.~A. Aguilar-Saavedra, 
{Nucl. Phys.} {\bf B813} (2009) 22--90,
  [arXiv:0808.2468 [hep-ph]].

  
\bibitem{Han:2006ip}
  T.~Han and B.~Zhang,
  Phys.\ Rev.\ Lett.\  {\bf 97} (2006) 171804
  [hep-ph/0604064]; 
  F.~del Aguila, J.~A.~Aguilar-Saavedra and R.~Pittau,
  JHEP {\bf 0710} (2007) 047
  [hep-ph/0703261]; 
  A.~Atre, T.~Han, S.~Pascoli and B.~Zhang,
  JHEP {\bf 0905} (2009) 030
  [arXiv:0901.3589 [hep-ph]].
  

\bibitem{Franceschini:2008pz}
  R.~Franceschini, T.~Hambye and A.~Strumia,
  Phys.\ Rev.\ D {\bf 78} (2008) 033002
  [arXiv:0805.1613 [hep-ph]]; 
  F.~del Aguila and J.~A.~Aguilar-Saavedra,
  Phys.\ Lett.\ B {\bf 672} (2009) 158
  [arXiv:0809.2096 [hep-ph]]; 
  A.~Arhrib, B.~Bajc, D.~K.~Ghosh, T.~Han, G.~-Y.~Huang, I.~Puljak and G.~Senjanovic,
  Phys.\ Rev.\ D {\bf 82} (2010) 053004
  [arXiv:0904.2390 [hep-ph]].
  
 
\bibitem{Chatrchyan:2012ya}
  S.~Chatrchyan {\it et al.}  [CMS Collaboration],
  Eur.\ Phys.\ J.\ C {\bf 72} (2012) 2189
  [arXiv:1207.2666 [hep-ex]].
 
\bibitem{ATLAS:2012hi}
  G.~Aad {\it et al.}  [ATLAS Collaboration],
  Eur.\ Phys.\ J.\ C {\bf 72} (2012) 2244
  [arXiv:1210.5070 [hep-ex]].
 
  
\bibitem{delAguila:2008pw}
  F.~del Aguila, J.~de Blas and M.~Perez-Victoria,
  {Phys. Rev.} {\bf D78} (2008) 013010
  [arXiv:0803.4008 [hep-ph]].


\bibitem{Dev:2013wba}
  P.~S.~B.~Dev, A.~Pilaftsis and U.~-k.~Yang,
  arXiv:1308.2209 [hep-ph].
 

\bibitem{Chatrchyan:2012fla}
  S.~Chatrchyan {\it et al.}  [CMS Collaboration],
  Phys.\ Lett.\ B {\bf 717} (2012) 109
  [arXiv:1207.6079 [hep-ex]].

\bibitem{ATLAS:2012yoa}
  [ATLAS Collaboration],
  ATLAS-CONF-2012-139.

  
\bibitem{LR}
  J.~C.~Pati and A.~Salam,
  Phys.\ Rev. {\bf D10} (1974) 275 
  [Erratum-ibid. {\bf D11} (1975) 703];
  R.~N.~Mohapatra and J.~C.~Pati,
  Phys.\ Rev. {\bf D11} (1975) 2558;
  G.~Senjanovic and R.~N.~Mohapatra,
  Phys.\ Rev. {\bf D12} (1975) 1502.

\bibitem{WR_N}
  S.~N.~Gninenko, M.~M.~Kirsanov, N.~V.~Krasnikov and V.~A.~Matveev,
  Phys.\ Atom.\ Nucl.\  {\bf 70} (2007) 441;
  F.~del Aguila, J.~A.~Aguilar-Saavedra and J.~de Blas,
  Acta Phys.\ Polon.\ B {\bf 40} (2009) 2901
  [arXiv:0910.2720 [hep-ph]];
  P.~S.~B.~Dev, C.~-H.~Lee and R.~N.~Mohapatra,
  arXiv:1309.0774 [hep-ph].
  
\bibitem{CMS:2012zv}
  S.~Chatrchyan {\it et al.}  [CMS Collaboration],
  Phys.\ Rev.\ Lett.\  {\bf 109} (2012) 261802
  [arXiv:1210.2402 [hep-ex]].


\bibitem{Babu:2009aq}
  K.~S.~Babu, S.~Nandi and Z.~Tavartkiladze,
  Phys.\ Rev.\ D {\bf 80} (2009) 071702
  [arXiv:0905.2710 [hep-ph]]; 
  G.~Bambhaniya, J.~Chakrabortty, S.~Goswami and P.~Konar,
  arXiv:1305.2795 [hep-ph].


     
\bibitem{Grifols:1989xe}
  J.~A.~Grifols, A.~Mendez and G.~A.~Schuler,
  Mod.\ Phys.\ Lett.\ A {\bf 4} (1989) 1485. 
  
  A.~Djouadi, J.~Kalinowski and P.~M.~Zerwas,
  Z.\ Phys.\ C {\bf 70} (1996) 435
  [hep-ph/9511342].

\bibitem{Perez:2008ha}
  P.~Fileviez Perez, T.~Han, G.~-y.~Huang, T.~Li and K.~Wang,
  Phys.\ Rev.\ D {\bf 78} (2008) 015018
  [arXiv:0805.3536 [hep-ph]].
  
\bibitem{Hisano:2013sn}
  J.~Hisano and K.~Tsumura,
  Phys.\ Rev.\ D {\bf 87} (2013) 5,  053004
  [arXiv:1301.6455 [hep-ph]].

  \bibitem{Raidal:2008jk}
  M.~Raidal, A.~van der Schaaf, I.~Bigi, M. L.~Mangano, Yannis K.~ Semertzidis and others,
  Eur.\ Phys.\ J.\ C {\bf 57} (2008) 13
  [arXiv:0801.1826 [hep-ph]]; 
  Pran Nath, Brent D.~Nelson, Hooman Davoudiasl, Bhaskar Dutta, Daniel Feldman and others,
  Nucl.\ Phys.\ Proc.\ Suppl. {\bf 200-202} (2010) 185


\bibitem{Aoki:2011pz}
  M.~Aoki, S.~Kanemura and K.~Yagyu,
  Phys.\ Rev.\ D {\bf 85} (2012) 055007
  [arXiv:1110.4625 [hep-ph]].
  

\bibitem{delAguila:2013yaa}
  F.~del Aguila, M.~Chala, A.~Santamaria and J.~Wudka,
  Phys.\ Lett.\ B {\bf 725} (2013) 310
  [arXiv:1305.3904 [hep-ph]].

\bibitem{delAguila:2013hla}
  F.~del Aguila, M.~Chala, A.~Santamaria and J.~Wudka,
  EPJ Web Conf.\  {\bf 60}, 17002 (2013)
  [arXiv:1307.0510 [hep-ph]];
  F.~del Aguila, M.~Chala, A.~Santamaria and J.~Wudka,
  Acta Phys.\ Polon.\ B {\bf 44} (2013) 11,  2139
  [arXiv:1311.2950 [hep-ph]].


\bibitem{Alloul:2013raa}
  A.~Alloul, M.~Frank, B.~Fuks and M.~R.~de Traubenberg,
  Phys.\ Rev.\ D {\bf 88} (2013) 075004
  [arXiv:1307.1711 [hep-ph]]; and references there in.

 
\bibitem{Babu:1988ki}
  K.~S.~Babu,
  Phys.\ Lett.\ B {\bf 203} (1988) 132.

  
  \bibitem{Gunion:1996pq}
  J.~F.~Gunion, C.~Loomis and K.~T.~Pitts,
  eConf C {\bf 960625} (1996) LTH096
  [hep-ph/9610237]; 
V.~Rentala, W.~Shepherd and S.~Su,
Phys.\ Rev.\ D {\bf 84} (2011) 035004
[arXiv:1105.1379 [hep-ph]]; 
  M.~Aoki, S.~Kanemura and K.~Yagyu,
  Phys.\ Lett.\ B {\bf 702} (2011) 355
   [Erratum-ibid.\ B {\bf 706} (2012) 495]
  [arXiv:1105.2075 [hep-ph]]; 
  K.~Yagyu,
  arXiv:1304.6338 [hep-ph].

\bibitem{Ren:2011mh}
  B.~Ren, K.~Tsumura and X.~-G.~He,
  Phys.\ Rev.\ D {\bf 84} (2011) 073004
  [arXiv:1107.5879 [hep-ph]].



\bibitem{Chiang:2012dk}
  C.~-W.~Chiang, T.~Nomura and K.~Tsumura,
  Phys.\ Rev.\ D {\bf 85} (2012) 095023
  [arXiv:1202.2014 [hep-ph]].


\bibitem{Akeroyd:2005gt}
  A.~G.~Akeroyd and M.~Aoki,
  Phys.\ Rev.\ D {\bf 72} (2005) 035011
  [hep-ph/0506176].


\bibitem{Han:2007bk}
  T.~Han, B.~Mukhopadhyaya, Z.~Si and K.~Wang,
  Phys.\ Rev.\ D {\bf 76} (2007) 075013
  [arXiv:0706.0441 [hep-ph]].


\bibitem{Williams:1934ad}
  E.~J.~Williams,
  Phys.\ Rev.\  {\bf 45} (1934) 729.
E.J. Williams, Phys.Rev. 45 (1934) 729(L); C.F. Weizs\"acker, Z.Phys. 88 (1934) 612.

\bibitem{Drees:1994zx}
  M.~Drees, R.~M.~Godbole, M.~Nowakowski and S.~D.~Rindani,
  Phys.\ Rev.\ D {\bf 50} (1994) 2335
  [hep-ph/9403368].


\bibitem{Gripaios:2009pe}
  B.~Gripaios, A.~Pomarol, F.~Riva and J.~Serra,
  JHEP {\bf 0904} (2009) 070
  [arXiv:0902.1483 [hep-ph]]; 
  J.~Mrazek, A.~Pomarol, R.~Rattazzi, M.~Redi, J.~Serra and A.~Wulzer,
  Nucl.\ Phys.\ B {\bf 853} (2011) 1
  [arXiv:1105.5403 [hep-ph]]; 
  E.~Bertuzzo, T.~S.~Ray, H.~de Sandes and C.~A.~Savoy,
  JHEP {\bf 1305} (2013) 153
  [arXiv:1206.2623 [hep-ph]];
  M.~Chala,
  JHEP {\bf 1301} (2013) 122
  [arXiv:1210.6208 [hep-ph]]; 
  L.~Vecchi,
  arXiv:1304.4579 [hep-ph].


\bibitem{Muhlleitner:2003me}
  M.~Muhlleitner and M.~Spira,
  Phys.\ Rev.\ D {\bf 68} (2003) 117701
  [hep-ph/0305288].
  

\bibitem{Alwall:2011uj}
  J.~Alwall, M.~Herquet, F.~Maltoni, O.~Mattelaer and T.~Stelzer,
  JHEP {\bf 1106} (2011) 128
  [arXiv:1106.0522 [hep-ph]].

\bibitem{Christensen:2008py}
  N.~D.~Christensen and C.~Duhr,
  Comput.\ Phys.\ Commun.\  {\bf 180} (2009) 1614
  [arXiv:0806.4194 [hep-ph]].


\bibitem{Sugiyama:2012yw}
  H.~Sugiyama, K.~Tsumura and H.~Yokoya,
  Phys.\ Lett.\ B {\bf 717} (2012) 229
  [arXiv:1207.0179 [hep-ph]].


\bibitem{Kanemura:2013vxa}
  S.~Kanemura, K.~Yagyu and H.~Yokoya,
  arXiv:1305.2383 [hep-ph]; 
  C.~Englert, E.~Re and M.~Spannowsky,
  arXiv:1306.6228 [hep-ph].


\bibitem{Sjostrand:2006za}
  T.~Sjostrand, S.~Mrenna and P.~Z.~Skands,
  JHEP {\bf 0605} (2006) 026
  [hep-ph/0603175].
  
\bibitem{Ovyn:2009tx}
  S.~Ovyn, X.~Rouby and V.~Lemaitre,
  arXiv:0903.2225 [hep-ph].

\bibitem{Cacciari:2011ma}
  M.~Cacciari, G.~P.~Salam and G.~Soyez,
  Eur.\ Phys.\ J.\ C {\bf 72} (2012) 1896
  [arXiv:1111.6097 [hep-ph]].
 
\bibitem{Conte:2012fm}
  E.~Conte, B.~Fuks and G.~Serret,
  Comput.\ Phys.\ Commun.\  {\bf 184} (2013) 222
  [arXiv:1206.1599 [hep-ph]].

\bibitem{Frederix:2011ss}
  R.~Frederix, S.~Frixione, V.~Hirschi, F.~Maltoni, R.~Pittau and P.~Torrielli,
  JHEP {\bf 1202} (2012) 099
  [arXiv:1110.4738 [hep-ph]].
     
\bibitem{Frixione:2002ik}
  S.~Frixione and B.~R.~Webber,
  JHEP {\bf 0206} (2002) 029
  [hep-ph/0204244].


\bibitem{Duerr:2013dza}
  M.~Duerr, P.~Fileviez Perez and M.~B.~Wise,
  Phys.\ Rev.\ Lett.\  {\bf 110} (2013) 231801
  [arXiv:1304.0576 [hep-ph]]; 
  P.~Fileviez Perez and M.~B.~Wise,
  Phys.\ Rev.\ D {\bf 88} (2013) 057703
  [arXiv:1307.6213 [hep-ph]].

 
\bibitem{Read:2002hq}
  A.~L.~Read,
  J.\ Phys.\ G {\bf 28} (2002) 2693.

\end{thebibliography}


\end{document}